\documentclass{article}

\usepackage{PRIMEarxiv}

\usepackage[utf8]{inputenc} 
\usepackage[T1]{fontenc}    
\usepackage{hyperref}       
\usepackage{url}            
\usepackage{booktabs}       
\usepackage{amsfonts}       
\usepackage{amsmath}
\usepackage{nicefrac}       
\usepackage{microtype}      
\usepackage{lipsum}
\usepackage{graphicx}
\usepackage{subcaption}
\usepackage{mathpazo}
\usepackage{sectsty}
\usepackage{tikz}
\usepackage[capitalize]{cleveref}
\usepackage[shortlabels,inline]{enumitem}
\usepackage{float}
\usepackage{csquotes}
\usepackage{tabularx}
\newcolumntype{L}{@{\extracolsep{\fill}}l}
\newcolumntype{R}{@{\extracolsep{\fill}}r}
\newcolumntype{C}{@{\extracolsep{\fill}}c}

\title{Bayesian dynamic mode decomposition for real-time ship motion digital twinning}

\author{
  Giorgio Palma$^{a}$, Andrea Serani$^{a}$, Kevin McTaggart$^{b}$, Shawn Aram$^{c}$, David W. Wundrow$^{c}$, \\ \textbf{David Drazen$^{c}$, Matteo Diez$^{a,\star}$}\\
  $^{a}$National Research Council-Institute of Marine Engineering, Rome, Italy\\
  $^{b}$Defence Research and Development Canada-Atlantic Research Center, Dartmouth, Nova Scotia, Canada\\
  $^{c}$Naval Surface Warfare Center-Carderock Division, West Bethesda, Maryland, U.S.A.\\
  $^\star$\texttt{matteo.diez@cnr.it} \\
}

\begin{document}

\begin{tikzpicture}[remember picture,overlay]
   \node [rectangle, fill=cyan, fill opacity=0.5, anchor=north, minimum width=\paperwidth, minimum height=3cm] at (current page.north) {};

   \node [anchor=north, minimum width=\paperwidth, minimum height=3cm, text width=\textwidth, align=center, text height=5ex, text depth=15ex, align=left] at (current page.north) {
     \sffamily\small
     \textbf{This is a preprint submitted to:} \textit{Ocean Engineering}
   };
\end{tikzpicture}

\maketitle

\begin{abstract}
Digital twins are widely considered enablers of groundbreaking changes in the development, operation, and maintenance of novel generations of products. They are meant to provide reliable and timely predictions to inform decisions 
along the entire product life cycle.
One of their most interesting applications in the naval field is the digital twinning of ship performances in waves, a crucial aspect in design and operation safety.
In this paper, a Bayesian extension of the Hankel dynamic mode decomposition method is proposed for ship motion's nowcasting as a prediction tool for naval digital twins.
The proposed algorithm meets all the requirements for formulations devoted to digital twinning, being able to adapt the resulting models with the data incoming from the physical system, using a limited amount of data, producing real-time predictions, and estimating their reliability. 
Results are presented and discussed for the course-keeping of the 5415M model in beam-quartering sea state 7 irregular waves at  
Fr = 0.33, using data from three different CFD solvers.
The results show predictions keeping good accuracy levels up to five wave encounter periods, with the Bayesian formulation improving the deterministic forecasts. 
In addition, a connection between the predicted uncertainty and prediction accuracy is found. 
\end{abstract}


\section{Introduction}\label{s:intro}
The seakeeping and maneuverability performance of ships and their structures' strength are critical to ensure the safety of the vessels' structures, payload, and crew. 
This is particularly true when thinking about severe sea conditions and adverse weather a mission may face, with climate change also playing a role in causing more frequent and extremely harsh operational environments.
Recent NATO Science and Technology Organization (STO) Applied Vehicle Technology (AVT) task groups focused on the development of prediction methods for ship seakeeping and maneuvering in waves (AVT-280 “Evaluation of Prediction Methods for Ship Performance in Heavy Weather”, 2017–2019, and AVT-348 “Assessment of Experiments and Prediction Methods for Naval Ships Maneuvering in Waves”, 2021–2023).
The availability of predictive models for ship performance in waves is crucial in several aspects and phases of the ship's lifecycle.
In the early design phase, they help develop robust and high-quality projects with reduced costs and time-to-market, meeting the International Maritime Organization (IMO) Guidelines and NATO Standardization Agreements (STANAG) for commercial and military ships, respectively.
The ability to predict ship motions in real seas may also be exploited to develop advanced controllers, such as model predictive control approaches, for improved ship handling \cite{CODESSEIRA2021, dong2023, xu2023, zhang2024}.
Real-time processing of operational ship data can also assist in aiding tactical judgments in ship handling to improve safe and cost-saving navigation \cite{bekker2019}.
Such predictive tools are also key enablers, for example, for accurate vessel fatigue damage assessment and forecasting, which are, in turn, fundamental for decision-making regarding structural inspections, maintenance planning, and potential ship life extension while keeping high safety standards \cite{vanderhorn2022}.
These examples motivate the relevance of developing a digital twin for ship motions in the maritime industry's digitalization process.

Digital twins are now widely recognized as catalysts for revolutionary changes in the development, operation, and maintenance of new-generation products. According to the definition in \cite{digitaltwin}, \textit{“a Digital Twin is a set of virtual information constructs that mimic the structure, context, and behavior of an individual/unique physical asset, is dynamically updated with data from its physical twin throughout its lifecycle, and informs decisions that realize value”}. By reflecting the status of physical entities, digital twins provide value through reliable and timely predictions that guide decisions across the entire product lifecycle.
There is a common but often misleading perception that digital twins are the highest-fidelity virtual representations of all aspects of the real system of interest. However, time and resource-efficiency requirements impose digital twins to be rather purpose-driven virtual models of a given physical system or process, and, hence, not unique assuming different forms depending on their purpose \cite{rasheed2020}. As \cite{box1979} famously stated, \textit{“all models are wrong but some are useful”}, the required accuracy level of the model and the acceptable computational cost are linked to the decision task the specific digital twin supports. This leads to the definition of a digital twin as \textit{“the best available models, sensor information, and input data to mirror and predict activities/performance over the life of its corresponding physical twin”} \cite{west2017}.

Some examples of digital twins applications to the ship domain can already be found in the literature.
In \cite{schirmann_ship_2018} a digital twin for ship motions is presented and used to help estimate structural fatigue due to wave response. 
\cite{danielsen2018digital} developed a digital twin of an autonomous vessel, able to reconstruct ship positioning, heading, and speed, along with the rpm of the engine, for real-time engine condition monitoring and vessel system identification.
\cite{bekker2019} reports the conceptual design of a monitoring and decision-aiding system for a polar research vessel using full-scale ship-board measurements in real time.
\cite{lee2022} presented a sophisticated digital twin for ship operation in seaways, including, among the others, computational modules for the reconstruction and prediction of ocean waves from marine radar images, and the simulation of the ship's seakeeping and maneuvering using data from GPS, MRU, and hydrodynamic database, feeding a deterministic equation-based model of the wave-induced ship motions.
Digital twin models are a versatile concept, whose characteristics respond to peculiar requirements of their specific purpose and context.
At the same time, the mandate of digital twins to evolve together with their real siblings 
introduces specific challenges and defines common requirements for formulations devoted to digital twinning. They have to be:
\begin{enumerate}[i)]
    \item adaptive in nature and conceived to continuously learn from data to represent the asset in its \enquote{as-is} state 
    (this is a key property that distinguishes them from the parent family of digital models);
    \item a close to real-time representation of the physical twin;
    \item able to learn predictive models from a limited amount of observations (\textit{i.e.,} to be data-lean as opposite to data-greedy);
    \item able to characterize the reliability of the estimated predictions.
\end{enumerate}

Considering these points, a first skim of the most promising existing approaches can be done. 
Computational fluid dynamics (CFD) methods, although able to predict ship performances including highly non-linear dynamics in extreme wave conditions \cite{serani2021urans,aram2024cfd}, are not suitable for real-time applications due to their typically very high computational costs (which is particularly true when complex hydro-structural problems are investigated via high-fidelity solvers, see \cite{lee2024}).
Semi-empirical reduced order models with physics-based simplifications and equation-based formulations can provide fast estimates, with a high level of accuracy, provided that they are properly calibrated \cite{WEEMS2023,LU2015}.
Nevertheless, they may lack the necessary adaptivity to follow the evolution of the physical system during its life (\textit{e.g.}, the engine degradation, or hull and propeller fouling).
On the other side, machine and 
learning methods are data-driven approaches that were shown to be extremely effective for forecasting ship motion variables \cite{Diez2024}. Their strength is the ability to capture hidden relationships and dynamics directly from available data, without prior knowledge of the system equations. Despite accurate and fast predictions can be obtained, these methods require a typically non-negligible data-intensive training effort, scarcely compliant with the point iii).

The dynamic mode decomposition (DMD, \cite{kutz2016dynamic}) is a data-driven, equation-free technique initially developed in fluid dynamics, where it has been used to identify coherent spatiotemporal structures and predict their evolution. Examples include vorticity generated by bluff bodies \cite{Tang2012}, jet turbulence \cite{Semeraro2012}, noise in jet mixing layers \cite{Song2013}, stability analyses of complex flows \cite{rowley2009, schmid2010} and magnetized plasma \cite{kaptanoglu2020}.
Originally introduced to identify linear normal modes in linear systems, DMD has since evolved to capture the underlying dynamics and coherent structures of complex, nonlinear systems by leveraging the Koopman operator concept.
DMD has found applications across various scientific and engineering fields, including structural dynamics, epidemiology \cite{Proctor2015}, and neuroscience \cite{brunton2016}, where it extracts spatio-temporal coherent patterns from large-scale neural recordings. 

DMD has been shown to be broadly applicable to the analysis of complex systems \cite{Brunton2021}, partly due to its data-driven nature. This characteristic makes DMD particularly well-suited for practical applications where obtaining a precise mathematical model is challenging or impractical.
The method, in fact, can operate on measured or simulated data without requiring any specific assumptions about system dynamics. Essentially, it identifies underlying structures directly from the data, bypassing the need for detailed system modeling.

In this paper, DMD is proposed as a nowcasting method that can be readily included in a digital twin dedicated to ship motions analysis. Specifically, the objective of the present work is to use the modes and frequencies identified by the DMD to build reduced-order predictive models (ROMs) of the motions of a ship navigating in irregular waves, capable of forecasting the system evolution at least in a short time window and respecting the requisites to be applied in digital twins.
The forecasting of ship trajectories/motions/forces by DMD has been previously addressed in \cite{diez2021datadriven,serani2024snh} and \cite{serani2023} with a statistical assessment of the algorithm forecasting capabilities, in \cite{diez2022} developing a hybrid prediction algorithm combining DMD and artificial neural network, and in \cite{Diez2024} comparing the performances of augmented state DMD and different neural network architectures.

In the current paper, a variant of the classic DMD that extends the state of the system under analysis with its time-delayed copies, called Hankel-DMD (also known as Augmented DMD or delay DMD) \cite{Brunton2017,kamb2020time,serani2023}, will be applied in an original Bayesian formulation that defines the prediction as a normally distributed uncertain variable characterized by mean value and standard deviation.
The algorithm performs the \textit{nowcasting} of the ship motion's variables, and the resulting models will comply with the four points addressed above, making them suitable for digital twin applications:
i) the forecasting algorithm will continuously update its approximation of the system ingesting new data, describing the ship in its ongoing evolution; 
ii) the prediction will be extremely fast to evaluate as the variables are expressed as a modal expansion, leading to a real-time forecast;
iii) the models will be obtained from relatively few snapshots from the near past of the variables to be predicted, with no need for huge training databases;
iv) the Bayesian formulation of the DMD algorithm is applied to introduce uncertainty in the estimated prediction.

Results are presented for the course keeping of a free-running naval destroyer (5415M) in irregular beam-quartering waves at Froude number Fr = 0.33 and sea state 7.
Data used to test the proposed algorithms are taken from physics-based numerical simulations conducted with three computational models, namely TEMPEST \cite{hughes2011tempest}, ShipMo3D \cite{McTaggart2019}, and CFDShip-Iowa \cite{huang2008-IJNMF}.

The structure of the paper is organized as follows. \Cref{s:meth} presents the DMD formulations, introducing the standard version, the Hankel-DMD, and, finally, the Hankel-DMD Bayesian extension used for the current study. 
A brief description of the test cases is given in \cref{s:testcases}, with some detail on the simulation tools that produced the dataset used in the current analysis.
\Cref{s:nset} describes the numerical setup for the DMD. 
\Cref{s:res} presents and discusses the numerical results, and, finally, conclusions on the use of DMD as a prediction tool for ships in waves in digital twins and possible extension for future work are discussed in \Cref{s:concl}.

\section{Methods}\label{s:meth}

\subsection{Dynamic mode decomposition}\label{s:dmd}
In general terms, the DMD can be interpreted as a technique for determining the eigenvalues and eigenvectors (modes) of a finite-dimensional linear model. This model serves as an approximation of the infinite-dimensional linear Koopman operator \cite{kutz2016dynamic}, also referred to as the composition operator, which is an infinite-dimensional linear operator that describes the time evolution of the system under study, even when nonlinearities are present.

A dynamic system can be modeled by:
\begin{equation}\label{eq:sysdyn}
\dfrac{\mathrm{d}\mathbf{x}}{\mathrm{d}t}=\mathbf{f}(\mathbf{x},t,\gamma),
\end{equation}
where $\mathbf{x}(t)\in\mathbb{R}^n$ is the system state at time $t$, 
$\gamma$ is the set of system parameters, 
and $\mathbf{f}(\cdot)$ represents the system dynamics, possibly including nonlinearities. 
\Cref{eq:sysdyn} can represent various types of systems, 
including the discretization of partial differential equations at some discrete spatial points, 
or the evolution of multi-variable time series.
DMD is an equation-free approach that returns a model of the system $\mathbf{f}(\mathbf{x},t,\gamma)$, that is considered unknown, without any assumption of the underlying physics. Once obtained, the model can be used to forecast the system behavior.
The method is based on observed data, such as time series of ship motions, and produces an approximation of the system model of \cref{eq:sysdyn} as a locally linear dynamical system \cite{kutz2016dynamic}:
\begin{equation}\label{eq:dmdsys}
\dfrac{\mathrm{d}\mathbf{x}}{\mathrm{d}t}=\mathcal{A}\mathbf{x},
\end{equation}
with solution
\begin{equation}\label{eq:dmdrec}
\mathbf{x}(t)=\sum_{k=1}^n \boldsymbol{\phi}_k \, q_k(t)=
\sum_{k=1}^n \boldsymbol{\phi}_k \, b_k \,\exp(\zeta_k t),
\end{equation}
where $\boldsymbol{\phi}_k$ and $\zeta_k$ are, respectively, the eigenvectors and the eigenvalues of the matrix $\mathcal{A}$, $q_k$ are the time-varying modal coordinates, and the coefficients $b_k$ are the modal coordinates of the initial condition $\mathbf{x}_0$ in the eigenvector basis, for which
\begin{equation}
\mathbf{b}=\boldsymbol{\Phi}^{-1}\mathbf{x}_0.
\end{equation}

The state of the system is typically measured at $m$ discrete time steps and can be expressed as $\mathbf{x}_j = \mathbf{x}(j\Delta t)$ with $j = 1,\dotsc,\,m$ and sampling time $\Delta t$. 
The dynamical system of \cref{eq:dmdsys} is hence conveniently replaced by its discrete-time version:
\begin{equation} 
    \mathbf{x}_{j+1} = \mathrm{\mathbf{F}}(\mathbf{x}_j,j,\gamma).
\end{equation}
Consequently, the DMD approximation can be written as
\begin{equation}\label{eq:dmddsys}
    \mathbf{x}_{j+1} = \mathrm{\mathbf{A}}\mathbf{x}_j, \hspace{1cm} \text{with} \hspace{1cm} \mathrm{\mathbf{A}}=\text{exp}(\mathcal{A}\Delta t).
\end{equation}
The matrix $\mathbf{A}$ in its DMD approximated version is defined as 
\begin{equation}\label{eq:approxA}
\mathbf{A}\approx\mathbf{X}'\mathbf{X}^{\dag},
\end{equation}
where $\mathbf{X}^{\dag}$ is the Moore-Penrose pseudo-inverse of $\mathbf{X}$, which minimizes $\|\mathbf{X}'-\mathbf{AX}\|_F$, where $\|\cdot\|_F$ is the Frobenius norm. 
The matrices $\mathbf{X}$ and $\mathbf{X}'$ are built arranging the $m$ system measurements as follows:
\begin{equation}\nonumber
\mathbf{X}=
\begin{bmatrix}
\mathbf{x}_j & \mathbf{x}_{j+1} & \dots & \mathbf{x}_{m-1}\\
\end{bmatrix},
\end{equation}
\begin{equation}\label{eq:XX'}
\mathbf{X}'=
\begin{bmatrix}
\mathbf{x}_{j+1} & \mathbf{x}_{j+2} & \dots & \mathbf{x}_{m}\\
\end{bmatrix},
\end{equation}
in which every column of $\mathbf{X}'$ is the snapshot of the system state a time step forward its homologous in $\mathbf{X}$.

In this work, the exact-DMD algorithm introduced by \cite{Tu2014} is applied. The pseudoinverse of $\mathrm{\mathbf{X}}$ can be efficiently evaluated using the singular value decomposition (SVD):
\begin{equation}\label{eq:svdapprox}
\mathrm{\mathbf{X}}^{\dagger} = \mathrm{\mathbf{V}} \boldsymbol{\Sigma}^{-1} \mathrm{\mathbf{U}}^*,    
\end{equation}
where $^*$ denotes the complex conjugate transpose. 
The matrix $\mathrm{\mathbf{\tilde A}}$ is evaluated by projecting $\mathrm{\mathbf{A}}$ 
onto the proper orthogonal decomposition modes in $\mathrm{\mathbf{U}}$ 
\begin{equation}
    \mathrm{\mathbf{\tilde A}} = \mathrm{\mathbf{U}}^* \mathrm{\mathbf{A}} \mathrm{\mathbf{U}},
\end{equation}
and its spectral decomposition can be evaluated
\begin{equation}
    \mathrm{\mathbf{\tilde A}} \mathrm{\mathbf{W}} = \mathrm{\mathbf{W}} \mathrm{\mathbf{\Lambda}}.
\end{equation}
The diagonal matrix $\mathrm{\mathbf{\Lambda}}$ contains the DMD eigenvalues $\lambda_k$, while the DMD 
eigenvectors
$\varphi_k$ constituting the matrix $\mathrm{\mathbf{\Phi}}$ are then reconstructed using the eigenvectors $\mathrm{\mathbf{W}}$ of the matrix $\mathbf{\tilde A}$ and the time-shifted data matrix $\mathrm{\mathbf{X}}'$ 
\begin{equation}\label{eq:modesdmd}
    \mathrm{\mathbf{\Phi}} = \mathrm{\mathbf{X}}' \mathrm{\mathbf{V}} \mathrm{\mathbf{\Sigma}}^{-1} \mathrm{\mathbf{W}}
\end{equation}
The time evolution of the state variables can be expressed by the following modal expansion, as per Eq. \ref{eq:dmdrec},
\begin{equation}\label{eq:dmdevol}
\mathbf{x}(t)=\sum_{k=1}^n \boldsymbol{\varphi}_k \, q_k(t)=
\sum_{k=1}^n \boldsymbol{\varphi}_k \, b_k \,\exp(\omega_k t),    
\end{equation}
where $\omega_k = \ln{(\lambda_k)}/\Delta t$.

Due to the low dimensionality of data in the current context, Eq. \ref{eq:svdapprox} is computed using the full SVD decomposition, with no rank truncation.

\subsection{Hankel dynamic mode decomposition}\label{s:dmdfore}
The base DMD formulation 
approximates the Koopman operator in a restricted space of linear measurements, creating a best-fit linear model linking sequential data snapshots \cite{schmid2010,kutz2016dynamic}. 
This linear DMD provides a locally linear representation of the dynamics that can't capture many essential features of nonlinear systems.
The augmentation of the system state with nonlinear functions of the measurements is thus the subject of several DMD algorithmic variants \cite{otto2019,takeishi2017,Lusch2018}, aiming to find a coordinate system (or \textit{embedding}) that spans a Koopman-invariant subspace.
However, there is no general rule for defining these observables and guaranteeing they will form a closed subspace under the Koopman operator \cite{brunton2016b}.

The Hankel-DMD \cite{Brunton2017,kamb2020time}
is a specific version of the DMD algorithm that has been developed to deal with the cases in which only partial observations of the system are available such that there are \textit{latent} variables.
Here, the state vector is thus augmented, embedding time-delayed copies of the original variables, resulting in an intrinsic coordinate system that forms a Koopman-invariant subspace (the time-delays form a set of observable functions that span a finite-dimensional subspace of Hilbert space, in which the Koopman operator preserves the structure of the system, see \cite{Brunton2017, Pan2020}).

The extension of base DMD formulation presented in \cref{s:dmd} is achieved by transforming the matrices $\bf X$ and ${\bf X}'$ in the augmented matrices $\widehat{\bf X}$ and $\widehat{\bf X}'$. Augmentation is achieved by adding a set of $s$ time-shifted copies of the original states (delayed states) to the data matrix \cite{kamb2020time}.
The delayed snapshots can be arranged in the form of Hankel matrices:  
\begin{equation}\label{eq:sts'}
\mathbf{S}=
\begin{bmatrix}
\mathbf{x}_{j-1} & \mathbf{x}_{j} & \dots & \mathbf{x}_{m-2}\\
\mathbf{x}_{j-2} & \mathbf{x}_{j-1} & \dots & \mathbf{x}_{m-3}\\
\vdots & \vdots & \vdots & \vdots \\
\mathbf{x}_{j-s-1} & \mathbf{x}_{j-s} & \dots & \mathbf{x}_{m-s-1}\\
\end{bmatrix}, \quad
\mathbf{S}'=
\begin{bmatrix}
\mathbf{x}_{j} & \mathbf{x}_{j+1} & \dots & \mathbf{x}_{m-1}\\
\mathbf{x}_{j-1} & \mathbf{x}_{j} & \dots & \mathbf{x}_{m-2}\\
\vdots & \vdots & \vdots & \vdots \\
\mathbf{x}_{j-s} & \mathbf{x}_{j-s+1} & \dots & \mathbf{x}_{m-s}\\
\end{bmatrix},
\end{equation}
thus, the augmented versions of the matrices in Eq. \ref{eq:XX'} are
\begin{equation}\label{eq:sXX'}
\widehat{\mathbf{X}}=
\begin{bmatrix}
\mathbf{X} \\
\mathbf{S}\\
\end{bmatrix},
\qquad
\widehat{\mathbf{X}}'=
\begin{bmatrix}
\mathbf{X}' \\ 
\mathbf{S}'\\
\end{bmatrix}.
\end{equation}
The procedure described above for the identification of the DMD eigenvalues and eigenvectors is, hence, applied with the only modification of the augmented matrices 
replacing the originals in \cref{eq:approxA} and subsequent equations.

The use of time-delayed copies as additional observables in the DMD has been connected to the Koopman operator as a universal linearizing basis \cite{Brunton2017}. 
DMD analyzes a set of time series by decomposing them into a corresponding set of modes, each with its own fixed frequency and growth or decay rate. This allows a highly complex problem with numerous states to be reduced to a relatively small number of modes. 
When the rank of the data matrix (typically the length of its column vectors) is small, time-delayed copies are useful for augmenting 
it and, consequently, 
the number of SVD modes that can be used for the dynamics modeling by the DMD, which can improve the accuracy of the prediction.

\subsection{Bayesian extension of Hankel dynamic mode decomposition}\label{s:bayDMD}
The shape and the entries of the matrix $\mathbf{A}$ built applying the Hankel-DMD
are influenced by the method's input hyper-parameters, 
such as the observation duration, $l_{tr} = t_m - t_1$, 
and the maximum delay duration $l_{d} = t_{j-1} - t_{j-s-1}$.
These dependencies can be denoted as follows:
\begin{equation}\label{eq:bayes1}
\mathbf{A}=\mathbf{A}(l_{tr},l_{d}).
\end{equation}
Through uncertainty propagation, the solution $\mathbf{x}(t)$ is also influenced by $l_{tr}$ and $l_d$:
\begin{equation}\label{eq:bayes2}
\mathbf{x}(t)=\mathbf{x}(t,l_{tr},l_{d}),
\end{equation}
In a Bayesian framework, the prediction 
provided by Hankel-DMD
is represented as 
a normally distributed stochastic variable with a mean value and standard deviation, defined, at a specific time $t$, by the following equations:
\begin{equation}\label{eq:bayes3}
\mathbf{\mu_x}(t)=\int_{l_{d}^l}^{l_{d}^u} \int_{l_{tr}^l}^{l_{tr}^u}\mathbf{x}(t,l_{tr},l_{d})p(l_{tr})p(l_{d})dl_{tr} dl_{d},
\end{equation}
\begin{equation}\label{eq:bayes4}
\mathbf{\sigma_x}(t)= \left\{ \int_{l_{d}^l}^{l_{d}^u} 
 \int_{l_{tr}^l}^{l_{tr}^u} \left[\mathbf{x}(t,l_{tr},l_{d})-\mathbf{\mu_x}(t) \right]^2 p(l_{tr})p(l_{d})dl_{tr} dl_{d} \right\}^\frac{1}{2},
\end{equation}
where ${l_{tr}^l}$, ${l_{d}^l}$ and ${l_{tr}^u}$, ${l_{d}^u}$ are lower and upper bounds and $p(l_{tr})$, $p(l_{d})$ are the given probability density functions for $l_{tr}$ and $l_{d}$.

In practice, a variation range $[{l_{tr}^l}, {l_{tr}^u}]$ for the observation time $l_{tr}$ and $[{l_{d}^l}, {l_{d}^u}]$ for the delay time $l_{d}$ is selected; both $l_{tr}$ and $l_{d}$ are assigned a uniform probability density.
Monte Carlo sampling is then applied and for each realization of $l_{tr} \in [{l_{tr}^l}, {l_{tr}^u}]$ and $l_{d} \in [{l_{d}^l}, {l_{d}^u}]$ the solution $\mathbf{x}(t,l_{tr},l_{d})$ is computed. At a given time $t$ the mean and standard deviation of the solutions are then evaluated.

\subsection{Performance metrics}\label{s:metrics}
To evaluate the predictions made by the models and to compare the effectiveness of different methodologies and configurations, three error indices are employed: the normalized mean square error (NRMSE) \cite{Diez2024}, the normalized average minimum/maximum absolute error (NAMMAE) \cite{Diez2024}, and the Jensen-Shannon divergence (JSD) \cite{marlantes2024}. 
All the metrics are averaged over the variables that constitute the system's state, providing a holistic assessment of prediction accuracy. This comprehensive evaluation considers aspects such as overall error, the range, and the correlation of predicted versus measured values.

The NRMSE quantifies the average root mean square error between the predicted values $\mathrm{\mathbf{\tilde x}}_t$ and the measured (test) values $\mathrm{\mathbf{x}}_t$ at different time steps. 
It is calculated by taking the square root of the average squared differences, normalized by the standard deviation of the measured values:
\begin{equation}\label{eq:nrmse}
   \mathrm{NRMSE} = \frac{1}{N} \sum_{i=1}^{N} \sqrt{\frac{1}{\mathcal{T} \sigma^2_{x_i}} \sum_{j=1}^{\mathcal{T}} \left( \tilde{x}_{ij} - x_{ij} \right)^2},
\end{equation}
where $N$ is the number of variables in the predicted state, $\mathcal{T}$ is the number of considered time instants, and $\sigma_{x_i}$ is the standard deviation of the measured values in the considered time window for the variable $x_i$.

The NAMMAE metric provides an engineering-oriented assessment of the prediction accuracy. It measures the absolute difference between the minimum and maximum values of the predicted and measured time series, as follows:
\begin{equation}
    \mathrm{NAMMAE} = \frac{1}{2 N \sigma_{x_i}} \sum_{i=1}^{N} \left( \left| \min_j(\tilde{x}_{ij}) - \min_j(x_{ij}) \right| + \left| \max_j(\tilde{x}_{ij}) - \max_j(x_{ij}) \right| \right),
\end{equation}

Lastly, the JSD provides a measure of the similarity between the probability distribution of the predicted and reference signal \cite{marlantes2024}.
For each variable, it estimates the entropy of the predicted time series probability density function $Q$ relative to the probability density function of the measured time series $R$, where $M$ is the average of the two \cite{Lin1991}. 
The Jensen-Shannon divergence is based on the Kullback-Leibler divergence $D$, given by \cref{eq:kld}, which is 
the expectation of the logarithmic difference between the probabilities $K$ and $H$, both defined over the domain $\chi$, where the expectation is taken using the probabilities $K$ \cite{Kullback1951}
\begin{align}
    &\mathrm{JSD} = \frac{1}{N} \sum_{i=1}^{N} \left( \frac{1}{2}D(Q_i\,||\,M_i) + \frac{1}{2}D(R_i\,||\,M_i) \right), \quad \text{with} \quad M = \frac{1}{2} (Q + R)\label{eq:jsd}\\
    &D(K\,||\,H)=\sum_{y \in \chi} K(y) \ln\left( \frac{K(y)}{H(y)} \right). \label{eq:kld}
\end{align}
The similarity between the distributions is higher when the Jensen-Shannon distance is closer to zero. JSD is upper bounded by $\ln(2)$.

Each of the three indices contributes to the error assessment with its peculiarity.
The NRMSE evidences phase, frequency, and amplitude errors between the reference and the predicted signal, evaluating a pointwise difference between the two. However, it is not possible to discern between the three types of error and to what extent each type is contributing to the overall value.
The NAMMAE indicates if the prediction varies in the same range of the original signal, but does not give any hint about the phase or frequency similarity of the two.
JSD index is ineffective in detecting phase errors between the predicted and the reference signals. Instead, it highlights if the compared time histories assume each value in their range of variation a similar number of times. It is hence sensible to prediction frequency and trend errors.
An example of synergic use of the three is given for the case of a prediction that immediately goes to zero.
In this case, the NRMSE has a subtle behavior that may mislead the interpretation of the results if used alone: with standardized data, a prediction that immediately goes to zero will produce an NRMSE close to the standard deviation of the observed signal (\textit{i.e.,} around 1). This may be lower, for example, than the error obtained with a prediction well capturing the trend of the observed time history but with a small phase shift, and be misleading on the real forecasting capability of the algorithm at hand. 
The introduction of NAMMAE and JSD helps to discriminate the mentioned situation since their values would instead be higher for a null prediction.

\section{Test case description}\label{s:testcases}
\begin{figure}[ht!]
\centering
\includegraphics[width=0.75\linewidth]{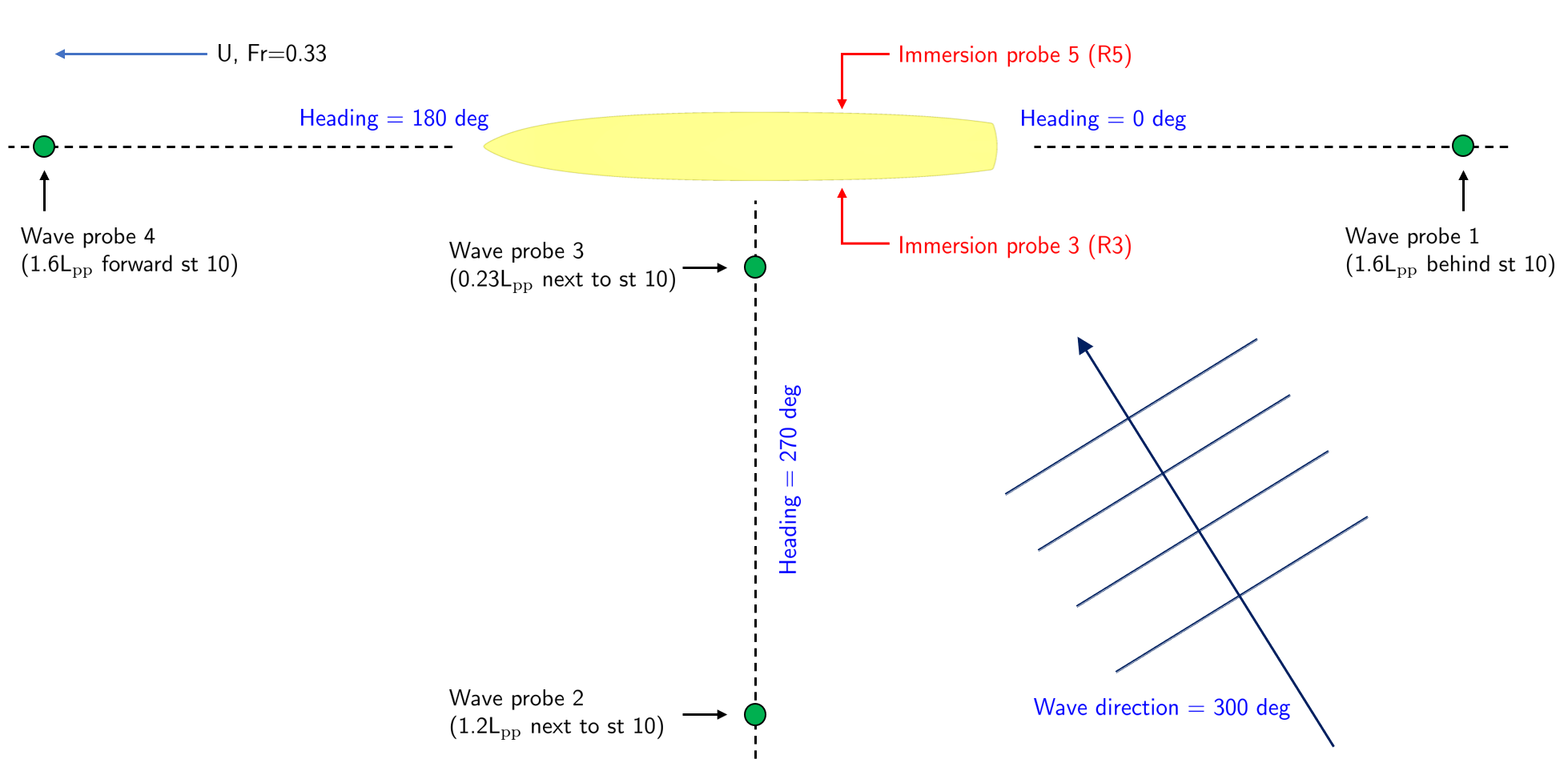}
  \caption{Setup and notation for the current test case \cite{serani2021urans}}
  \label{fig:scheme}
  \bigskip
\end{figure}

The hull form under investigation is referred to as 5415M, a modified form of the destroyer model DTMB 5415 that has been the subject of many model tests and numerical studies, such as the NATO
AVT-280 
\cite{WalreeVisser2010}. 
The present work considers the motions of the 
5415M in sea state 7 (high) according to the World Meteorological Organization definition, with a nominal significant wave height of 7 m. 
The simulations are conducted in irregular long-crested waves, with a nominal peak period $T_p$ =9.2 s and wave heading of 300 deg, i.e., beam-quartering seas (from 30 degrees aft of the beam, see \cref{fig:scheme}) and a nominal 
Fr = 0.33.
The combination of sea state, heading, and forward speed was selected to achieve a condition close to roll resonance and motions with significant nonlinearities.  

Course-keeping computations are based on the URANS code CFDShip-Iowa V4.5 \cite{huang2008-IJNMF}, and the potential flow codes TEMPEST \cite{hughes2011tempest} and ShipMo3D \cite{McTaggart2019}. 
In the simulations, the ship is kept on course by a simulated proportional derivative control actuating the rudder angle.

\begin{figure}[!ht]
    \centering
    \includegraphics[width=0.85\linewidth]{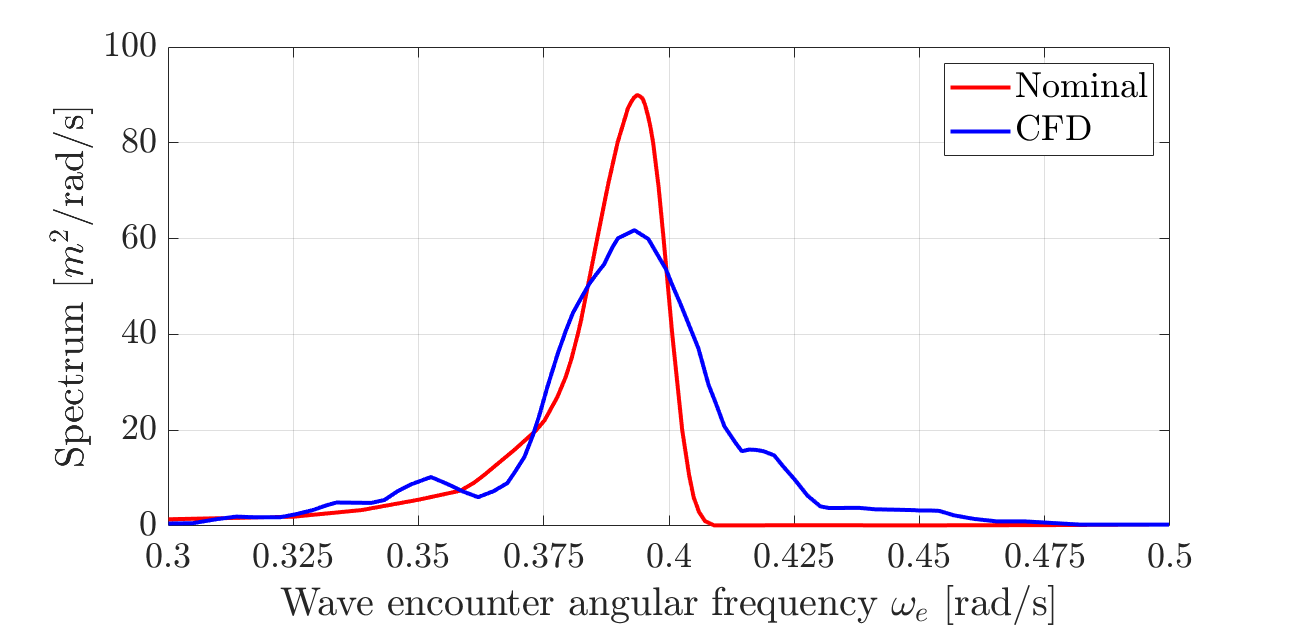}
    \caption{CFD versus nominal JONSWAP spectrum (encounter wave). Note that wave probe 2 is used for CFD analyses.}\label{fig:spectrum}
\end{figure}

CFDShip-Iowa was specifically developed for ship applications, 
including the prediction of ship maneuvering in calm water and waves.  
The current study uses RANS computations with isotropic implementation of Menter’s blended 
$k-\varepsilon/k-\omega$ (BKW) model with shear stress transport (SST) using no wall function. 
The six degrees of freedom rigid body equations of motion are solved to calculate the linear and angular motions of the ship.
The ship propeller is simulated using a simplified body-force model, which prescribes axisymmetric body force with axial and tangential components.  
Simulations in calm water were used to determine the constant propeller RPM for simulations in waves. 

TEMPEST was developed by the Naval Surface Warfare Center Carderock Division (NSWCCD) for the simulation of large amplitude ship motions, including capsize \cite{BelknapReed2019}.
TEMPEST uses potential flow modeling to evaluate radiation, diffraction, and incident wave forces on the hull \cite{bandyk2009body}.  
Nonlinear hydrostatic and wave excitation forces are evaluated based on the instantaneous wetted surface.  
Radiation and diffraction forces are evaluated using a body-exact strip-theory method.  
Like many ship motion codes based on potential flow, TEMPEST uses coefficient-based models for forces heavily influenced by viscous effects. Maneuvering forces are modeled using Abkowitz-type stability derivatives \cite{abkowitz1964ship}, and cross-flow lift and drag coefficients that vary with drift angle.  
Input hull maneuvering force terms for 5415M model were evaluated using double-body RANS simulations from NavyFOAM, which is an extension of OpenFOAM for naval hydrodynamics applications \cite{AramField2016,AramKim2017,Kim2017,Aram2018}.  
For propeller forces, TEMPEST applies the widely-used approach of considering thrust coefficients as functions of advance coefficients, which has been implemented using look-up tables with data from \cite{serani2021urans} for model 5415M.  
The influence of the hull on propeller inflow is modeled using straightening coefficients, which were evaluated for the destroyer using double-body RANS simulations.  
Rudder lift forces are evaluated using empirical models from \cite{whicker1958free}, and rudder inflow straightening was assessed using the same double-body RANS simulations used for propellers.

ShipMo3D was developed by Defence Research and Development Canada (DRDC) for the prediction of ship motions in frequency and time domains \cite{McTaggart2019}. 
ShipMo3D assumes potential flow when evaluating radiation, diffraction, and incident wave forces on the hull.  
The 3D panel method using the zero speed Green function evaluates radiation and diffraction forces in the frequency domain based on the calm waterline \cite{McTaggart2015}.
Forward speed effects are modeled to the extent possible, providing robust computations and acceptable accuracy for monohull ships with forward speed Froude numbers less than 0.5.
Like TEMPEST and many other ship motion codes, ShipMo3D uses coefficient-based methods for various forces terms, enabling it to capture phenomena such as viscous effects.
ShipMo3D input maneuvering coefficients were based on data from CFD computations \cite{sadat2015cfd}.  
Propeller RPM values for simulations were based on an iterative approach determining the required RPM in calm water for a given ship speed.
Rudder forces were evaluated using lift and drag coefficients based on input rudder geometry.  
Rudder inflow evaluation considers various phenomena, including flow straightening by the hull and outflow from the propellers.  

Each irregular seaway was modeled using linear superposition of 100 wave components, that were used to model a JONSWAP spectrum.  
The frequencies of the wave components $\omega_i$ were evenly distributed between 0.41 and 1.47 rad/s, with a constant frequency increment $\Delta \omega_i$ of 0.135 rad/s.  
The amplitude of each wave component was $\zeta_i = \sqrt{2 \: S(\omega_i) \: \Delta \omega_i}$.    
The phase of each wave component was randomly generated from a range between 0 and $2 \pi$.

\section{DMD numerical setup}\label{s:nset}
The CFDShip-Iowa data included 72,057 time steps covering 212 encountered waves from 8 simulation runs;
the TEMPEST data were from 49 simulations spanning 294,000 time steps over approximately 1000 encountered waves;
finally, the ShipMo3D data were from a single simulation with 25,001 time steps over 175 encountered waves. 
To provide a common baseline for processing by DMD, data from all the software were downsampled to 32 time steps per nominal wave encounter.
All DMD analyses are based on normalized data. Specifically, the Z-score is used; therefore time histories with zero mean and unit variance are evaluated through the process.

The state vector $\bf x$ used in the DMD analyses is composed of the ship's heave $x_3$, the three rigid body rotation roll $\phi$, pitch $\theta$, and yaw $\psi$, the rudder position angle $\alpha$, and the surge and sway velocities $v_1$ and $v_2$, respectively. 
It may be noted that here the DMD is proposed for the forecasting of ship motion data in a data-driven modeling problem, using a reduced set of variables that can be reasonably measured on board. 

Borrowing some terminology from the machine learning field, the portion of the data used in \cref{eq:XX'} and \cref{eq:sts'} to build the ROMs is referred to as the training set or training data. 
Similarly, the portion of time histories used to evaluate the performance of the DMD predictions is called the test set or test data.

\begin{figure}[ht!]
    \centering
    \includegraphics[width=\textwidth]{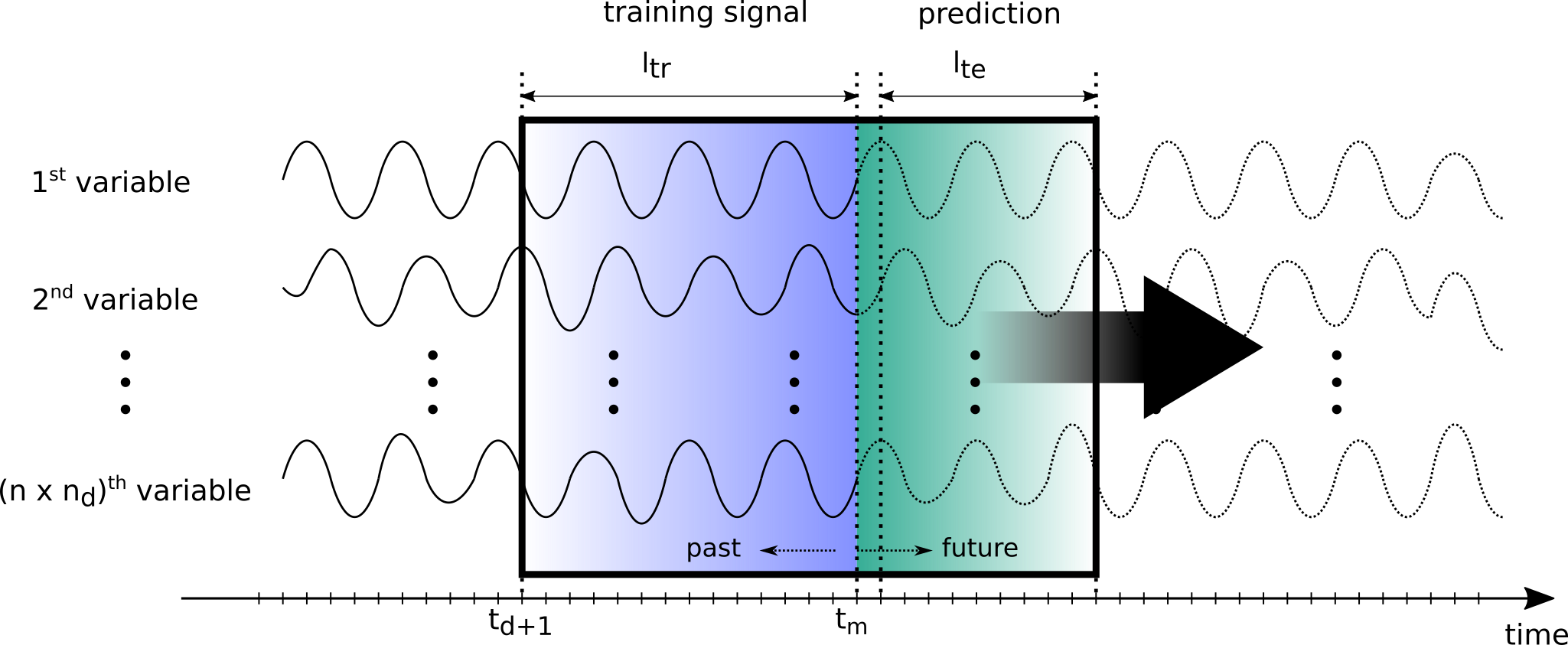}
    \caption{Sketch of the \textit{nowcasting} approach with Hankel-DMD.}
    \label{fig:sketch_nowcasting}
\end{figure}

The \textit{nowcasting} algorithm is based on the Hankel-DMD (and its Bayesian extension), and its main characteristic is that it creates a different ROM each time a prediction is required, using the near-past history of the variables. 
In other words, for any prediction starting point the matrices $\hat{\bf X}$ and $\hat{\bf X}'$ are built using the last $m$ known system measurements, a set of DMD modes and frequencies is then obtained, and 
the state evolution in time is finally predicted with \cref{eq:dmdevol}.
A sketch describing this approach is depicted in \cref{fig:sketch_nowcasting}.

The DMD in its Hankel variation (with and without control) exposes two main hyperparameters, \textit{i.e.}, the length of the training time histories, $l_{tr}$, and the number of delayed copies, that can be expressed as the shift of the most delayed embedded time history $l_d$. 
These two values deeply affect the prediction capability of the methods, while no general rule is given for the determination of their optimal values.
The Bayesian extension of the Hankel-DMD tries to overcome this issue, also introducing some information about the reliability of the estimated forecast which is obtained in terms of mean value and standard deviation.
As described in \cref{s:bayDMD}, the values of the hyperparameters are considered stochastic variables, and their ranges of variation are defined after a full-factorial numerical experiment in order to identify reasonable promising combinations. 

The influence of the hyperparameters on the forecasting capability of the \textit{nowcasting} algorithm is explored using 6 levels of variation, as reported in \cref{tab:doe_nc}, where $\hat T$ is a reference period, and $n_{tr}$ and $n_d$ are the observation and delay lengths respectively expressed in signal samples and number of additional states. The value of $\hat T$ is identified from data as the average value of the wave encounter periods on wave probe 3 signal.
\begin{table}[ht!]
    \caption{List of the hyperparameter tested settings for Hankel-DMD nowcasting algorithm}\label{tab:doe_nc}
    \centering
    \begin{tabular*}{0.5\linewidth}{@{} LLLLLLLL@{}}
     \toprule
      $l_{tr}$ & [-] &  $0.5\hat{T}$           & $\hat{T}$   & $2\hat{T}$ & $3\hat{T}$ & $4\hat{T}$ & $5\hat{T}$ \\
      $l_{d}$  & [-] & $0.5\hat{T}$           & $\hat{T}$   & $2\hat{T}$ & $3\hat{T}$ & $4\hat{T}$ & $5\hat{T}$ \\  
            \midrule
      $n_{tr}$ & [-] & 16 & 32 & 64 & 96 & 128 & 160 \\
      $n_{d}$  & [-] & 16 & 32 & 64 & 96 & 128 & 160 \\
      \bottomrule
    \end{tabular*}
\end{table}
 
All the analyses are done considering a prediction window that extends for five wave periods, $l_{te}\le 5\hat{T}$.

Each combination of the hyperparameters is tested on 250 different prediction starting points, statistically assessing the prediction capability of the setup.

Once the ranges of variation of the stochastic hyperparameters are defined, 
100 Monte Carlo realizations are considered for each time series to determine the expected value of the prediction and its associated standard deviation.

\section{Results and discussion}\label{s:res}
The \textit{nowcasting} algorithm employs the Hankel-DMD to predict the system’s state evolution from a specific initial time point. 
This method uses a relatively short portion of the near past time series up to the current time step to construct a linear ROM, defined by DMD modes and frequencies.
This ROM approximates the potentially non-linear system, keeping local accuracy over time but becoming less precise as the system progresses. 
To ensure continuous accurate predictions, a new ROM can be generated by repeating the process with a new starting point.

\subsection{Deterministic analysis}
\begin{figure}[ht!]
    \centering
    \includegraphics[width=\linewidth]{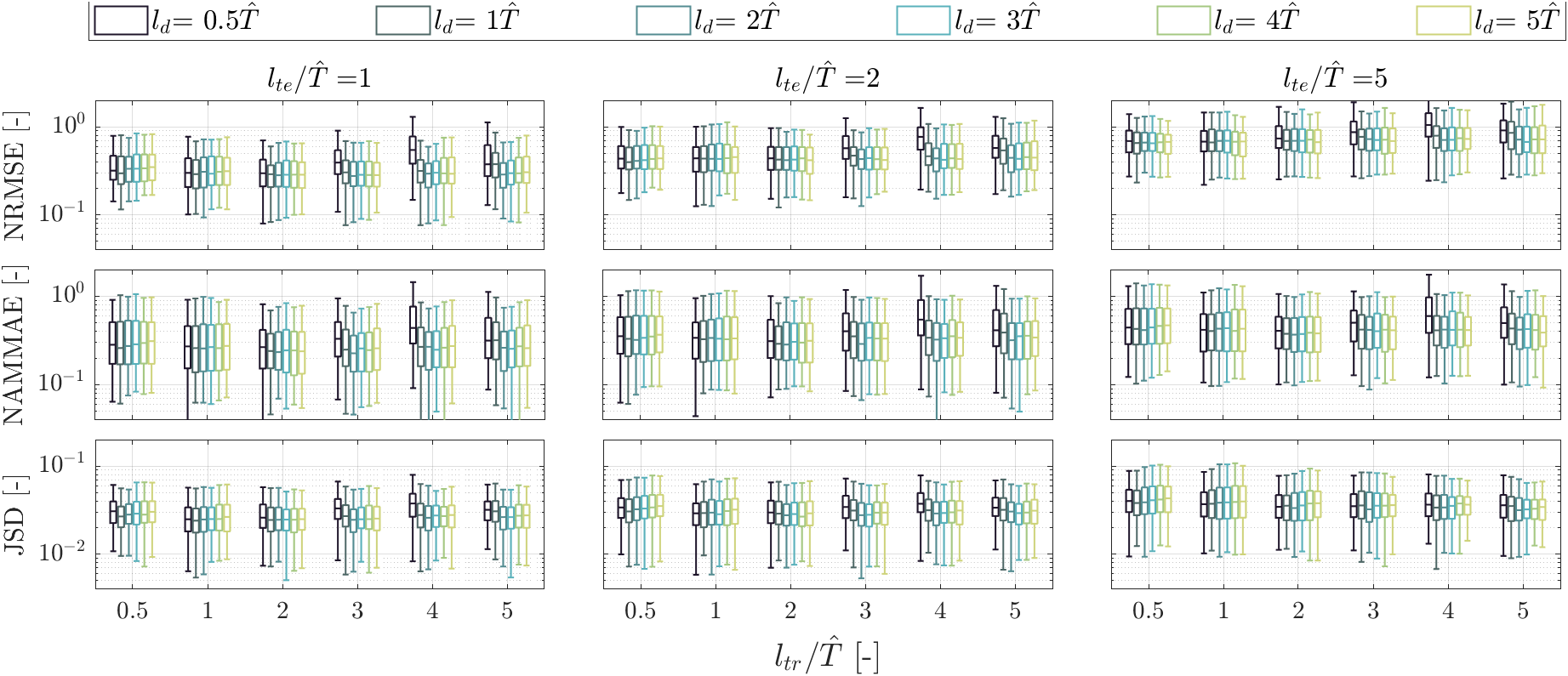}
    \caption{Error metrics from prediction statistical analysis of hyperparameters configurations, Hankel-DMD \textit{nowcasting} algorithm on CFDShip-Iowa data.}
    \label{fig:stat-iowa}
\end{figure}

\begin{figure}[ht!]
    \centering
    \includegraphics[width=\linewidth]{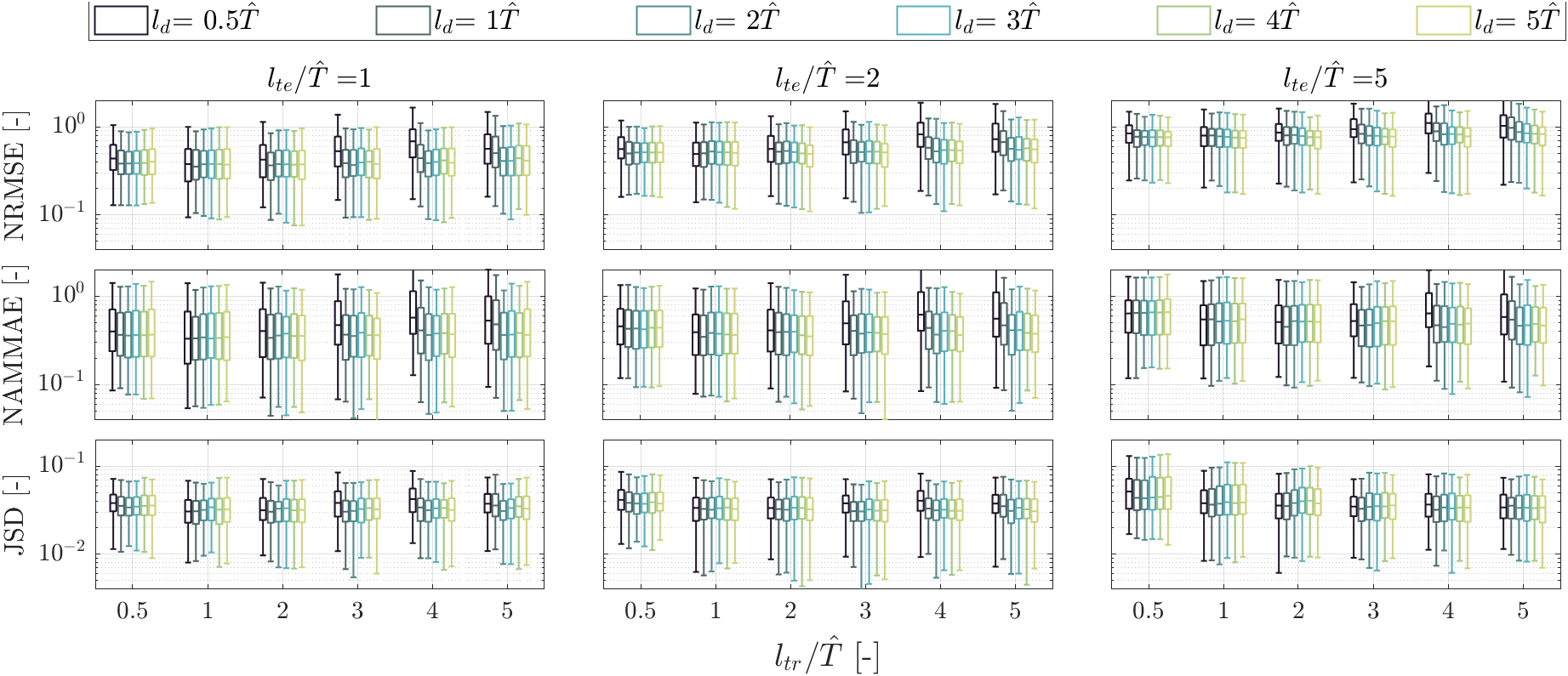}
    \caption{Error metrics from prediction statistical analysis of hyperparameters configurations, Hankel-DMD \textit{nowcasting} algorithm on TEMPEST data.}
    \label{fig:stat-tempest}
\end{figure}

\begin{figure}[ht!]
    \centering
    \includegraphics[width=\linewidth]{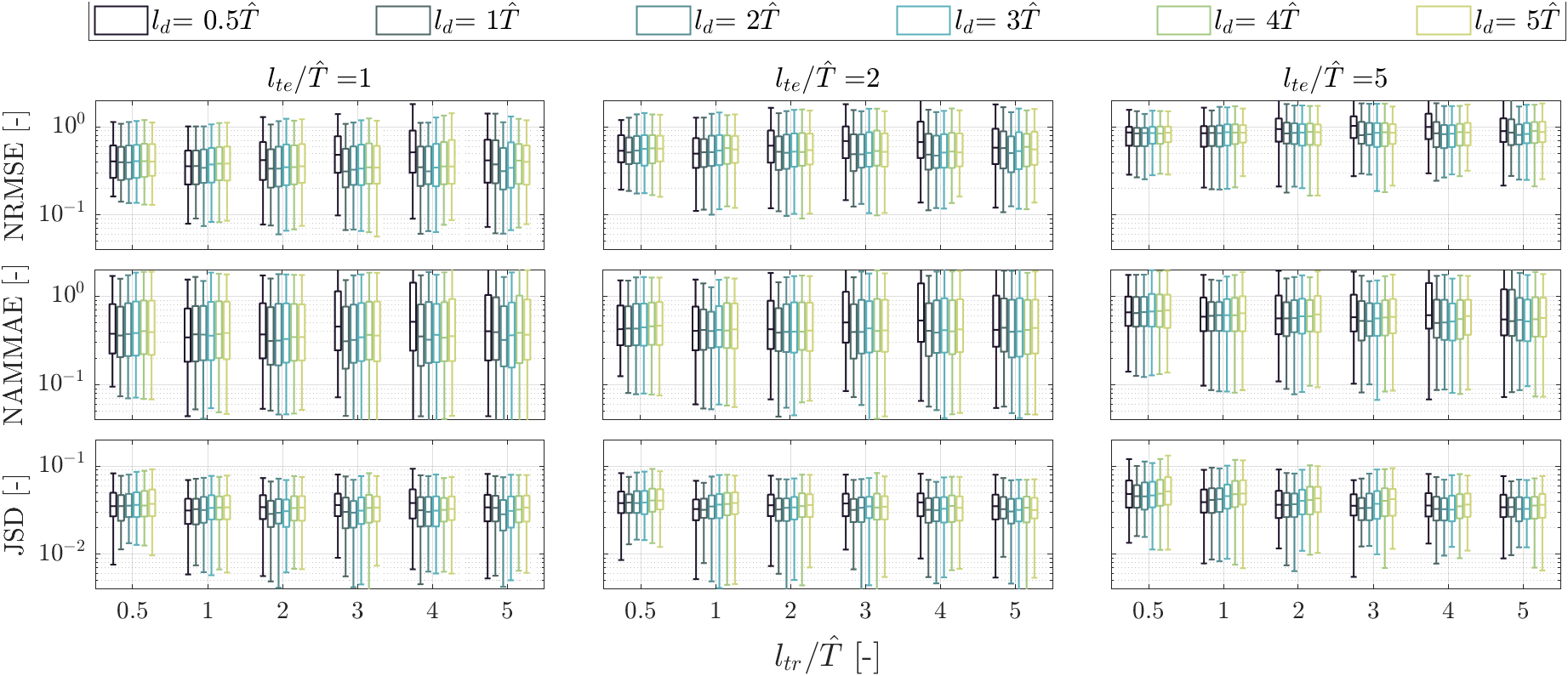}
    \caption{Error metrics from prediction statistical analysis of hyperparameters configurations, Hankel-DMD \textit{nowcasting} algorithm on ShipMo3D data.}
    \label{fig:stat-shipmo}
\end{figure}

\Cref{fig:stat-iowa,fig:stat-tempest,fig:stat-shipmo} report the results of the statistical analysis obtained using the deterministic version of the Hankel-DMD on the data from the three CFD solvers. 
The prediction capability of the algorithm is assessed through the performance metrics described in \cref{s:metrics}. 
Their values are presented as boxplots for each hyperparameter combination that has been tested. 
The boxes show the first, second (equivalent to the median value), and third quartiles, while the whiskers extend from the box to the farthest data point lying within 1.5 times the inter-quartile range, defined as the difference between the third and the first quartiles from the box. Outliers are not shown to improve the readability of the plot.
The values of the error metrics are presented for three different forecasting lengths $l_{te} = 1\hat{T}$, $2\hat{T}$, and $5\hat{T}$, respectively, showing the evolution of the prediction errors.
The results are also resumed in \cref{tab:det_an1,tab:det_an2,tab:det_an3}, for the same forecasting windows, in terms of mean value and standard deviation of the three error indices. In particular, we reported the metrics obtained with the hyperparameters configuration obtaining the best NRMSE average performance for each solver, \textit{i.e.}, $l_{tr}/\hat{T}=1$, $l_d/\hat{T}= 1$ for CFDShip-Iowa and TEMPEST, $l_{tr}/\hat{T}=4$, $l_d/\hat{T}= 2$ for ShipMo3D.
\begin{table}
\caption{Mean value and standard deviation of NRMSE for the Hankel-DMD in the best hyperparameter configuration and the Bayesian Hankel-DMD mean solution for each solver. Best configuration is $l_{tr}/\hat{T}=1$, $l_d/\hat{T}= 1$ for CFDShip-Iowa and TEMPEST, $l_{tr}/\hat{T}=4$, $l_d/\hat{T}= 2$ for ShipMo3D.}\label{tab:det_an1}
\begin{tabular*}{0.97\linewidth}{@{} LLLLLLL@{}}
      & NRMSE          &        &      & $l_{te}=\hat{T}$ & $l_{te}=2\hat{T}$ & $l_{te}=5\hat{T}$ \\
      \midrule
      & CFDShip-Iowa & Best deterministic     & avg    & 0.3329    & 0.4697     & 0.7417     \\
      &              &          & (std)  & (0.1684)  & (0.2101)   & (0.2950)   \\
      &              & Bayesian & avg    & 0.2736    & 0.4061     & 0.6626     \\
      &              &          & (std)  & (0.1222)  & (0.1730)   & (0.2628)   \\
      & TEMPEST      & Best deterministic     & avg    & 0.4247    & 0.5284     & 0.7940     \\
      &              &          & (std)  & (0.2422)  & (0.2312)   & (0.2882)   \\
      &              & Bayesian & avg    & 0.3744    & 0.5029     & 0.7653     \\
      &              &          & (std)  & (0.1900)  & (0.2237)   & (0.2878)   \\      
      & ShipMo3D     & Best deterministic     & avg    & 0.4683    & 0.6161     & 0.9223     \\
      &              &          & (std)  & (0.4018)  & (0.3958)   & (0.5283)   \\      
      &              & Bayesian & avg    & 0.4066    & 0.5398     & 0.8045     \\
      &              &          & (std)  & (0.3443)  & (0.3316)   & (0.3365)   \\
      \midrule
\end{tabular*}
\end{table}

\begin{table}
\caption{Mean value and standard deviation of NAMMAAE for the Hankel-DMD in the best hyperparameter configuration and the Bayesian Hankel-DMD mean solution for each solver. Best configuration is $l_{tr}/\hat{T}=1$, $l_d/\hat{T}= 1$ for CFDShip-Iowa and TEMPEST, $l_{tr}/\hat{T}=4$, $l_d/\hat{T}= 2$ for ShipMo3D.}\label{tab:det_an2}
\begin{tabular*}{0.97\linewidth}{@{} LLLLLLL@{}}
      & NAMMAE             &        &      & $l_{te}=\hat{T}$ & $l_{te}=2\hat{T}$ & $l_{te}=5\hat{T}$ \\
      \midrule
& CFDShip-Iowa & Best deterministic     & avg    & 0.2568    & 0.3423     & 0.4664     \\
      &              &          & (std)  & (0.1430)  & (0.2113)   & (0.2943)   \\
      &              & Bayesian & avg    & 0.2740    & 0.3447     & 0.4088     \\
      &              &          & (std)  & (0.2281)  & (0.2486)   & (0.2301)   \\
      & TEMPEST      & Best deterministic     & avg    & 0.5601    & 0.4964     & 0.6120     \\
      &              &          & (std)  & (0.7316)  & (0.4479)   & (0.4235)   \\
      &              & Bayesian & avg    & 0.4631    & 0.4737     & 0.5068     \\
      &              &          & (std)  & (0.5345)  & (0.3592)   & (0.3110)   \\
      & ShipMo3D     & Best deterministic     & avg    & 0.8262    & 0.7056     & 1.0695     \\
      &              &          & (std)  & (1.5377)  & (0.8280)   & (2.9190)   \\
      &              & Bayesian & avg    & 0.6812    & 0.5470     & 0.6296     \\
      &              &          & (std)  & (1.5124)  & (0.4970)   & (0.4840)   \\
\end{tabular*}
\end{table}

\begin{table}
\caption{Mean value and standard deviation of JSD for the Hankel-DMD in the best hyperparameter configuration and the Bayesian Hankel-DMD mean solution for each solver. Best configuration is $l_{tr}/\hat{T}=1$, $l_d/\hat{T}= 1$ for CFDShip-Iowa and TEMPEST, $l_{tr}/\hat{T}=4$, $l_d/\hat{T}= 2$ for ShipMo3D.}\label{tab:det_an3}
\begin{tabular*}{0.97\linewidth}{@{} LLLLLLL@{}}
      &  JSD            &        &      & $l_{te}=\hat{T}$ & $l_{te}=2\hat{T}$ & $l_{te}=5\hat{T}$ \\
      \midrule
   & CFDShip-Iowa & Best deterministic     & avg    & 0.0141    & 0.0217     & 0.0353     \\
      &              &          & (std)  & (0.0096)  & (0.0144)   & (0.0206)   \\
      &              & Bayesian & avg    & 0.0250    & 0.0293     & 0.0375     \\
      &              &          & (std)  & (0.0106)  & (0.0118)   & (0.0172)   \\
      & TEMPEST      & Best deterministic     & avg    & 0.0324    & 0.0348     & 0.0439     \\
      &              &          & (std)  & (0.0139)  & (0.0153)   & (0.0243)   \\
      &              & Bayesian & avg    & 0.0295    & 0.0308     & 0.0369     \\
      &              &          & (std)  & (0.0123)  & (0.0133)   & (0.0165)   \\
      & ShipMo3D     & Best deterministic     & avg    & 0.0333    & 0.0343     & 0.0351     \\
      &              &          & (std)  & (0.0158)  & (0.0135)   & (0.0147)   \\
      &              & Bayesian & avg    & 0.0308    & 0.0321     & 0.0356     \\
      &              &          & (std)  & (0.0160)  & (0.0140)   & (0.0154)   \\
\end{tabular*}
\end{table}
Observing \cref{fig:stat-iowa,fig:stat-tempest,fig:stat-shipmo}, a good consistency of the error trends is found among the three solvers.
It is noted that a high ratio between the observation length and the delay length (\textit{i.e.}, few additional delayed states and long training histories) results in increased errors. 
However, increasing the number of delayed time histories is not indefinitely beneficial; this is evident as, for a constant $l_{tr}$ and increasing $l_d$, the median lines of the box plots rise again after reaching a minimum.
Each observation length reveals the existence of an optimal range for the maximum delay time, which tends to increase with longer $l_{tr}$. 
By examining all three metrics and prediction windows, the optimal hyperparameter combinations can be claimed to lay in the ranges: 
\begin{equation}
    1 \le \frac{l_{tr}}{\hat T} \le 5, \hspace{1cm} \frac{1}{2} \le \frac{l_d}{l_{tr}} \le \frac{3}{4}.
\end{equation}

\begin{figure}[ht!]
    \centering
    \captionsetup[subfigure]{justification=centering}    
    \begin{subfigure}[b]{0.32\linewidth}       
        \includegraphics[width=\linewidth]{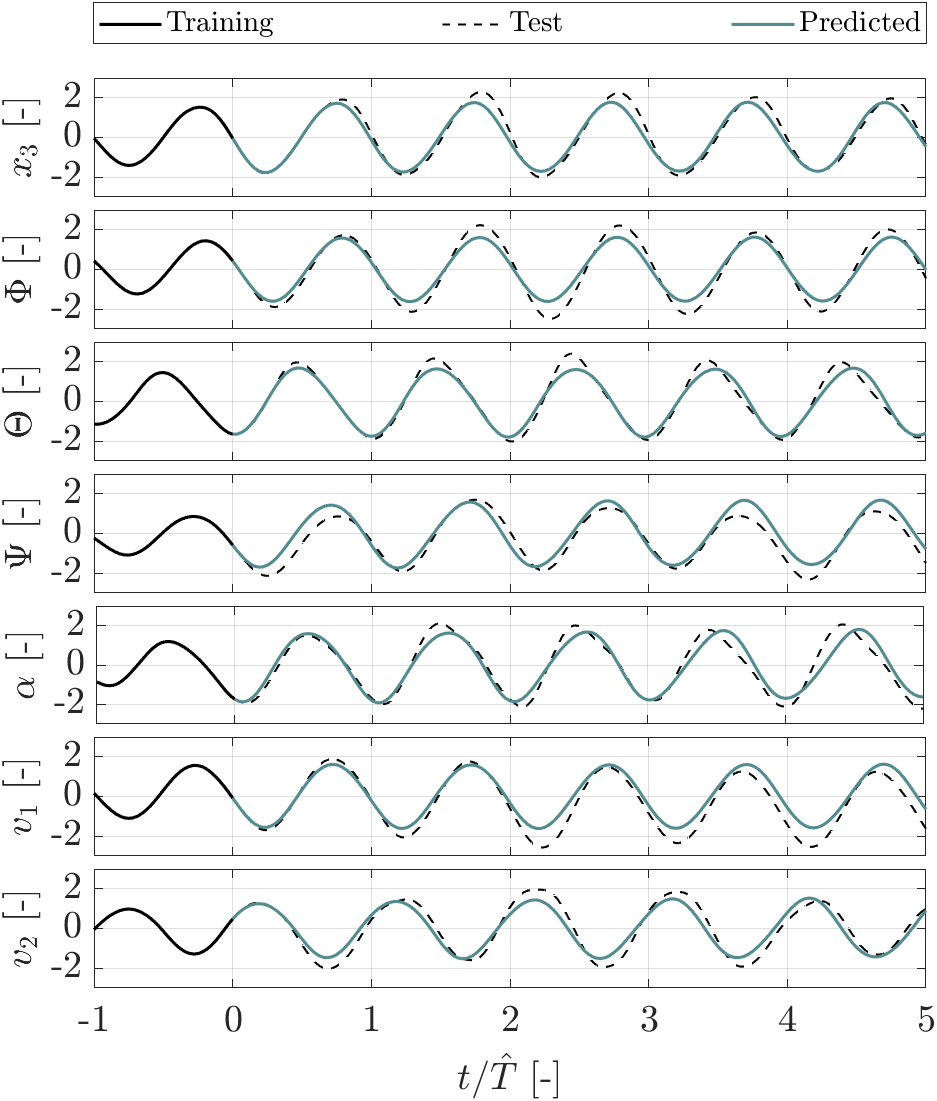}
        \caption{} \label{fig:fore1-iowa}
    \end{subfigure}
    \begin{subfigure}[b]{0.32\linewidth} 
        \includegraphics[width=\linewidth]{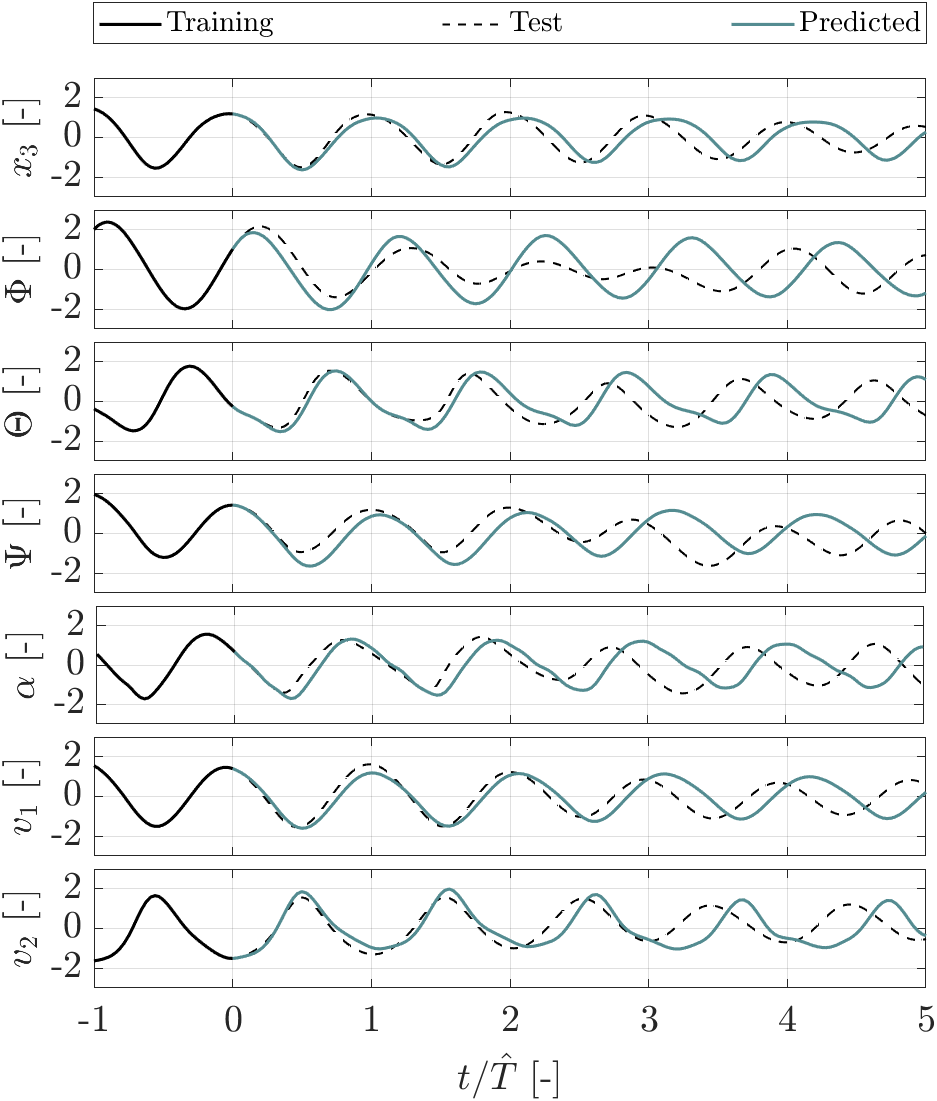}
        \caption{} \label{fig:fore1-tempest}    
    \end{subfigure}
    \begin{subfigure}[b]{0.32\linewidth}   
        \includegraphics[width=\linewidth]{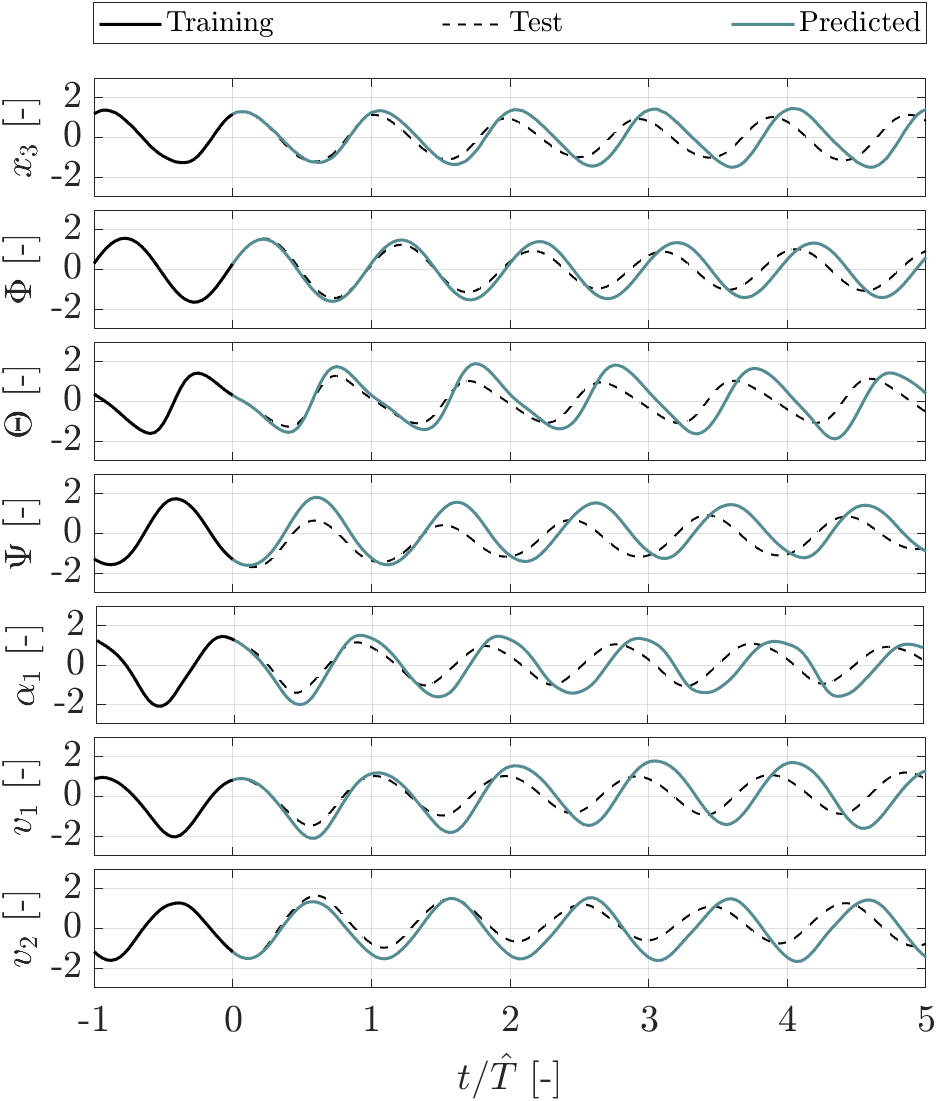}
        \caption{} \label{fig:fore1-shipmo3d}    
    \end{subfigure}
    \begin{subfigure}[b]{0.32\linewidth}       
        \includegraphics[width=\linewidth]{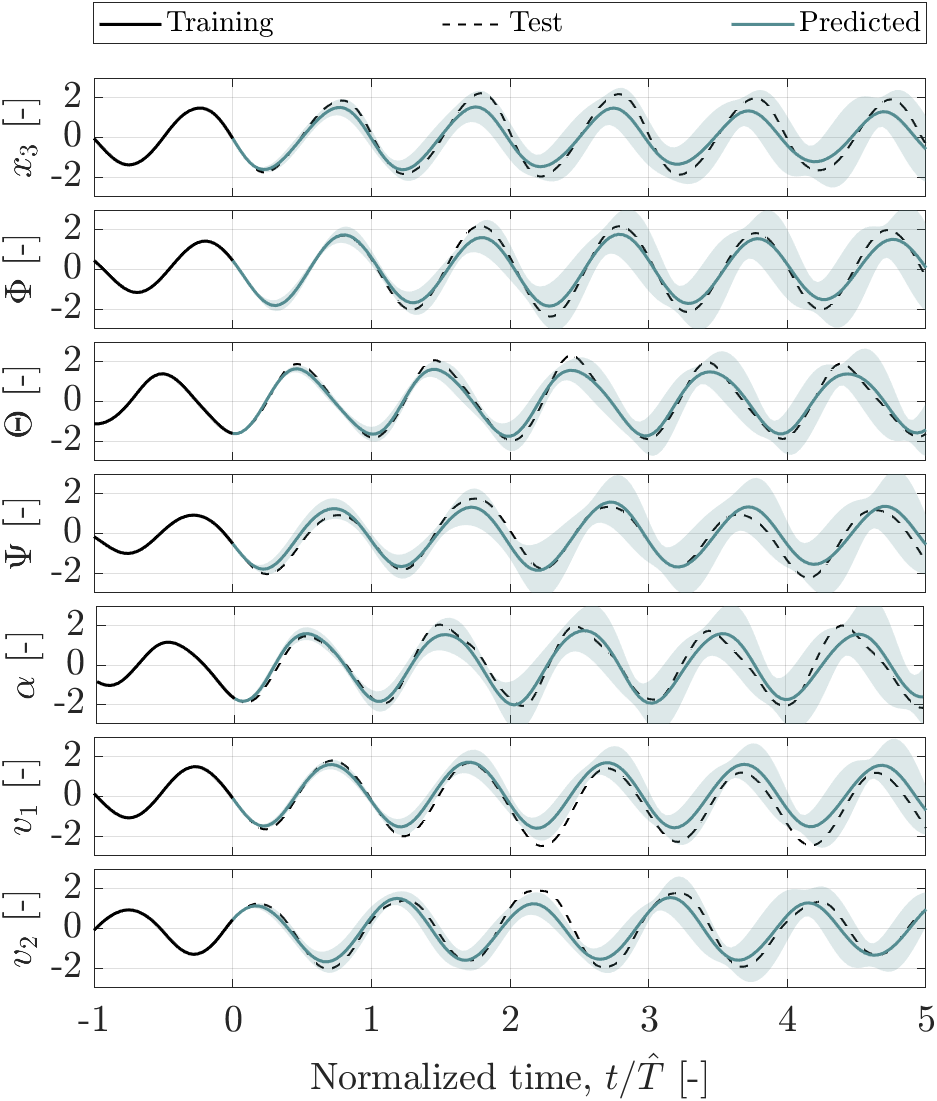}
        \caption{} \label{fig:bfore1-iowa}
    \end{subfigure}
    \begin{subfigure}[b]{0.32\linewidth} 
        \includegraphics[width=\linewidth]{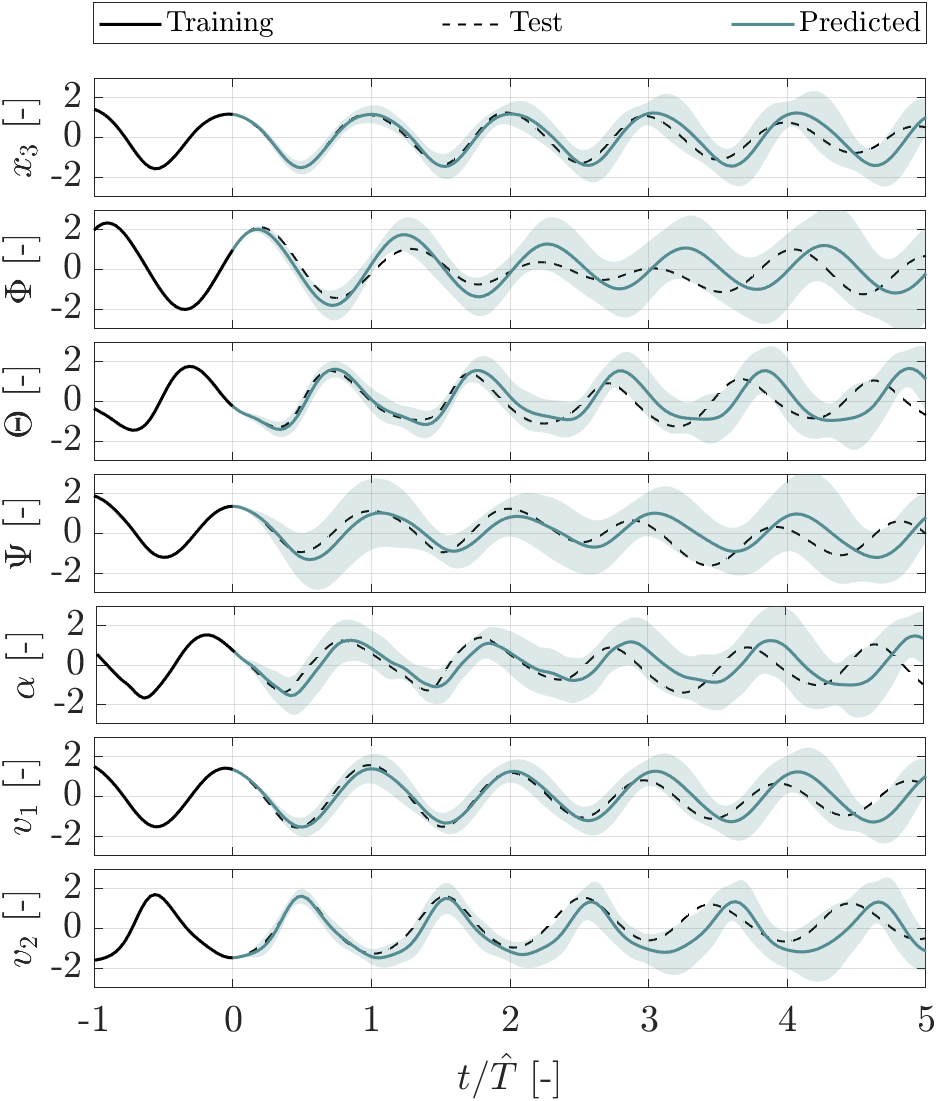}
        \caption{} \label{fig:bfore1-tempest}    
    \end{subfigure}
    \begin{subfigure}[b]{0.32\linewidth}   
        \includegraphics[width=\linewidth]{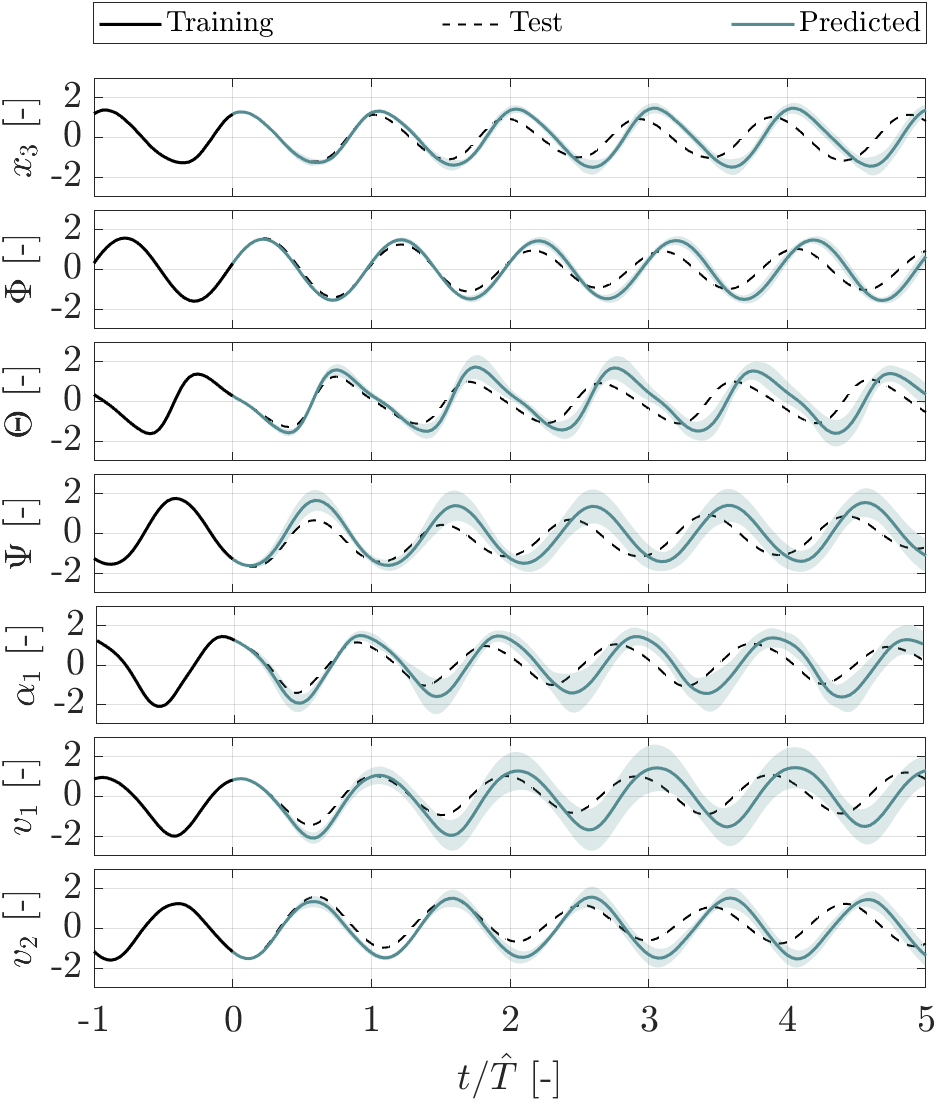}
        \caption{} \label{fig:bfore1-shipmo3d}    
    \end{subfigure}
\caption{\textit{Nowcasting} prediction of random time series by Hankel-DMD (best hyperparameters) and Bayesian Hankel-DMD. CFDShip-Iowa: (\subref{fig:fore1-iowa}) and (\subref{fig:bfore1-iowa}); TEMPEST: (\subref{fig:fore1-tempest}) and (\subref{fig:bfore1-tempest}); ShipMo3D: (\subref{fig:fore1-shipmo3d}) and (\subref{fig:bfore1-shipmo3d}).} \label{fig:fore1}
\end{figure}

\begin{figure}[ht!]
    \centering
    \captionsetup[subfigure]{justification=centering}   
    \begin{subfigure}[b]{0.32\linewidth}       
        \includegraphics[width=\linewidth]{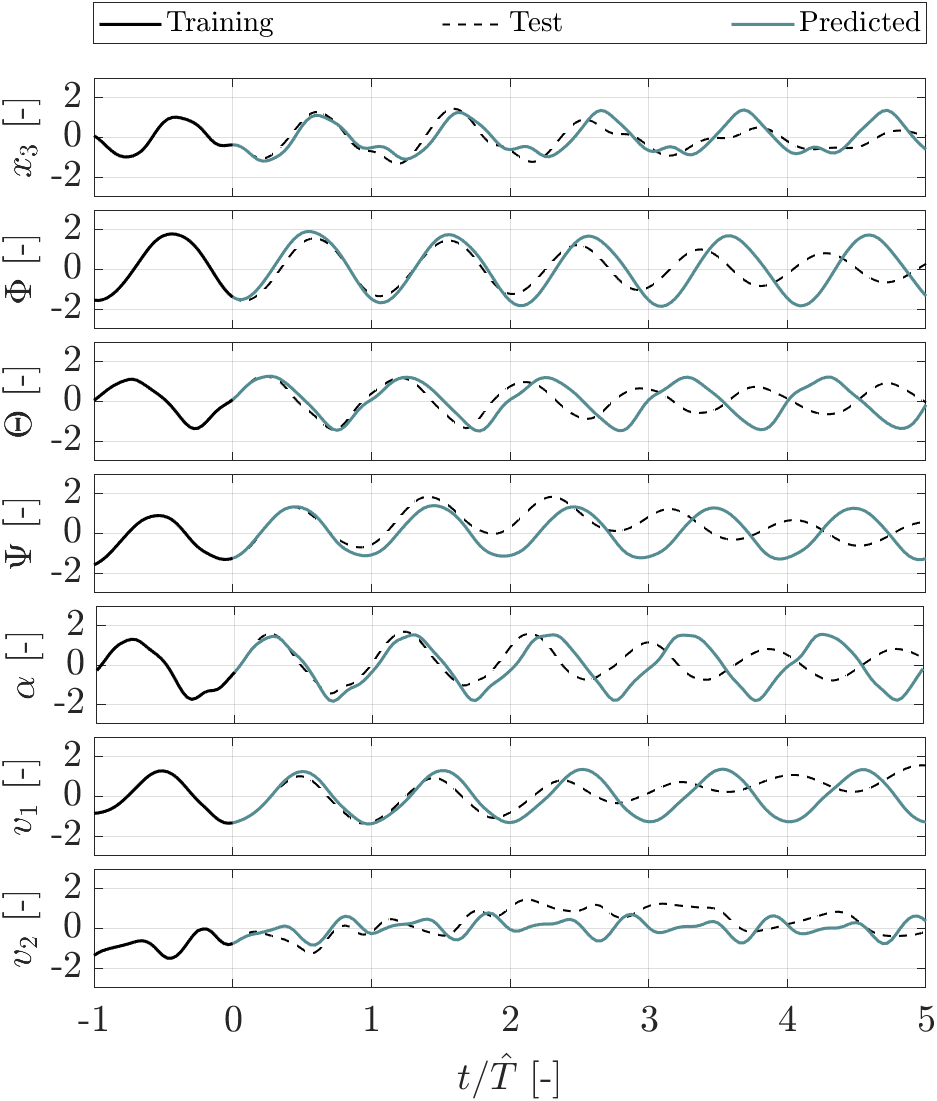}
        \caption{} \label{fig:fore2-iowa}
    \end{subfigure}
    \begin{subfigure}[b]{0.32\linewidth} 
        \includegraphics[width=\linewidth]{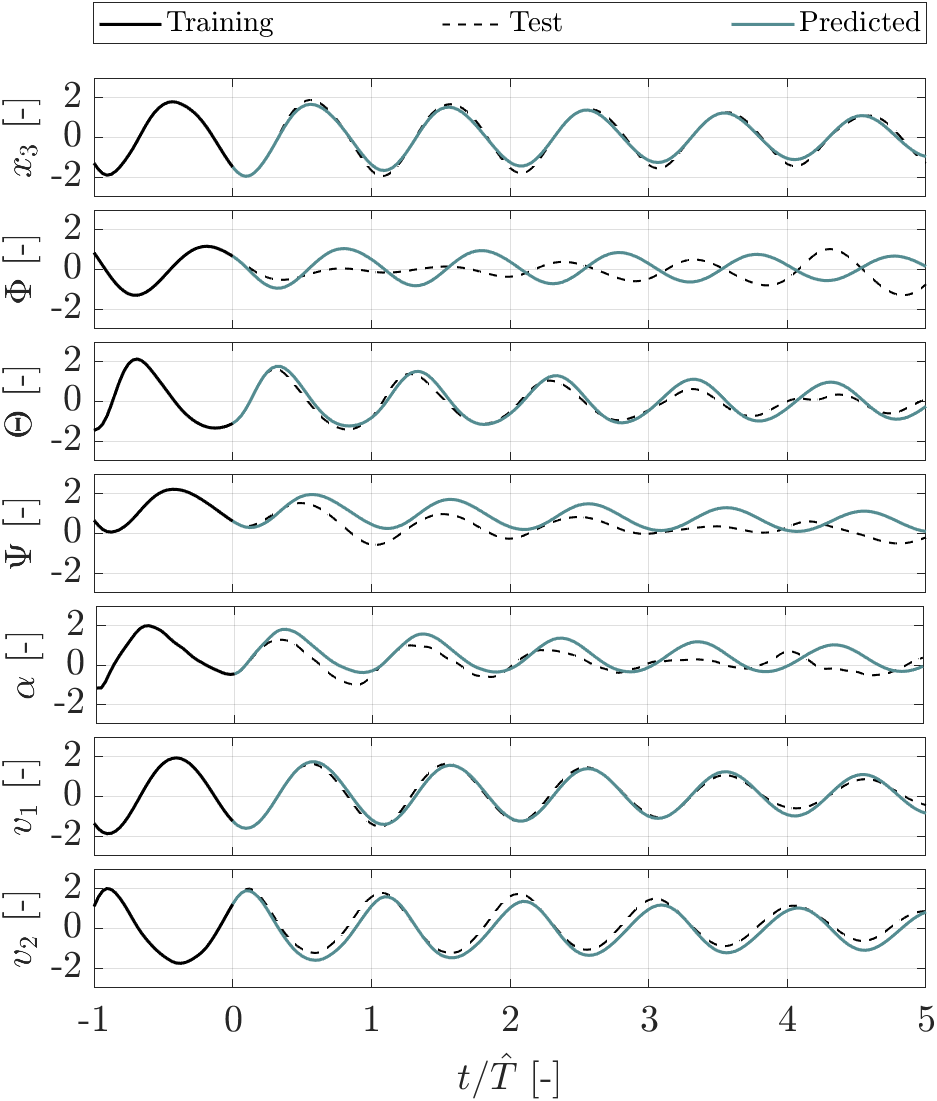}
        \caption{} \label{fig:fore2-tempest}    
    \end{subfigure}
    \begin{subfigure}[b]{0.32\linewidth}   
        \includegraphics[width=\linewidth]{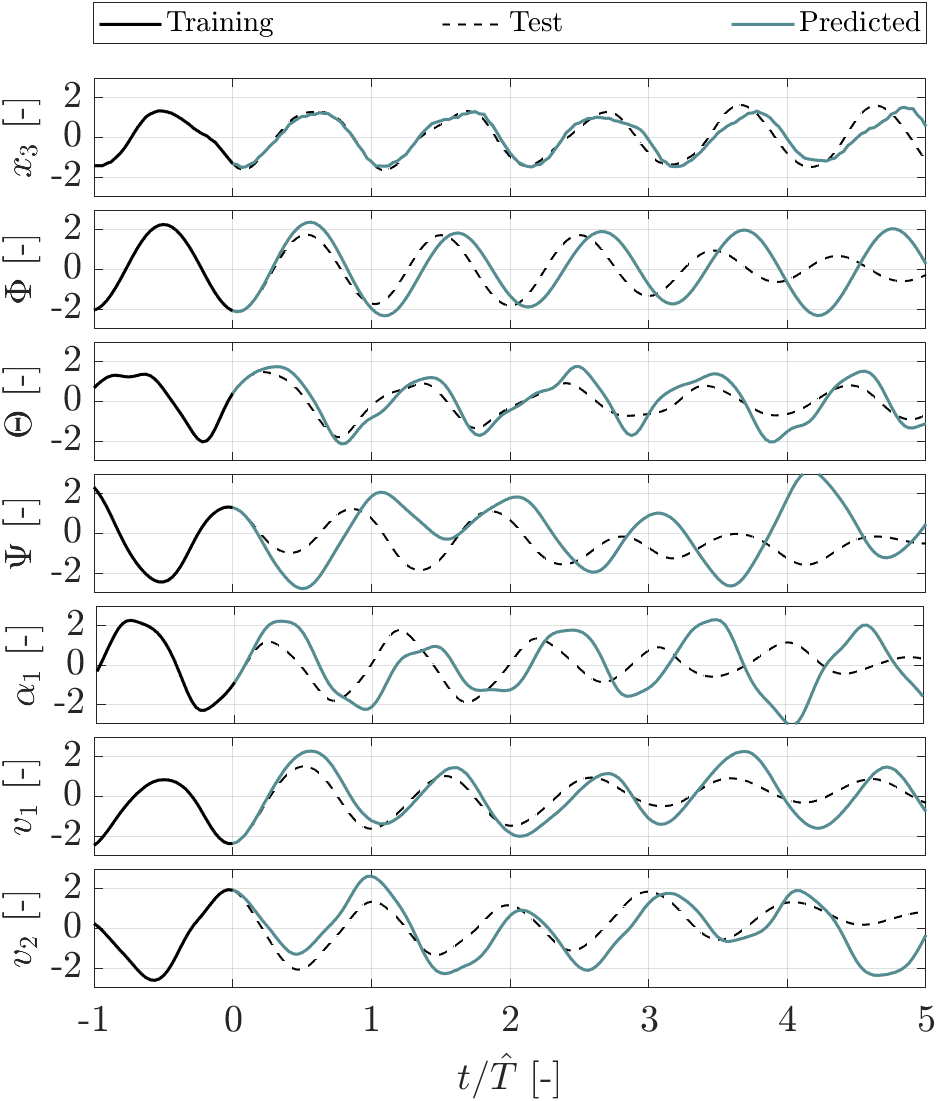}
        \caption{} \label{fig:fore2-shipmo3d}    
    \end{subfigure}
    \begin{subfigure}[b]{0.32\linewidth}       
        \includegraphics[width=\linewidth]{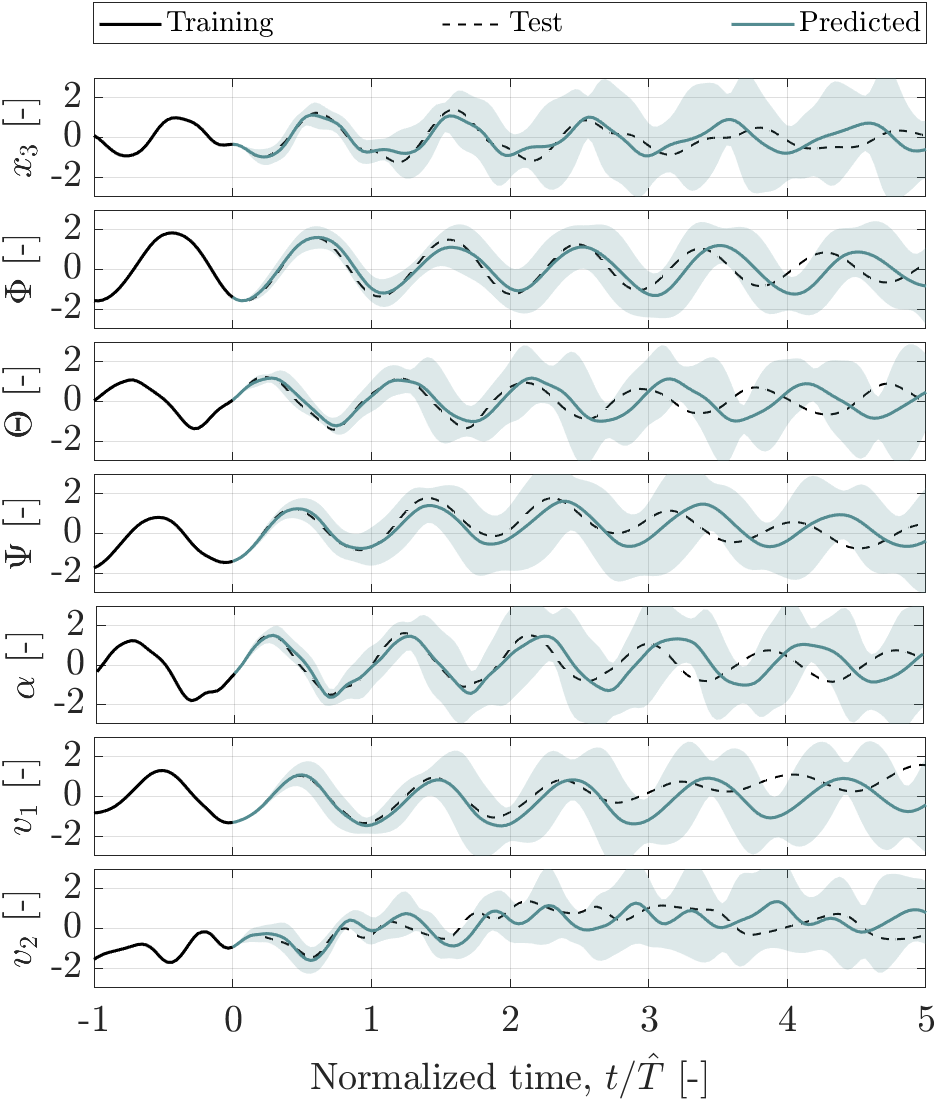}
        \caption{} \label{fig:bfore2-iowa}
    \end{subfigure}
    \begin{subfigure}[b]{0.32\linewidth} 
        \includegraphics[width=\linewidth]{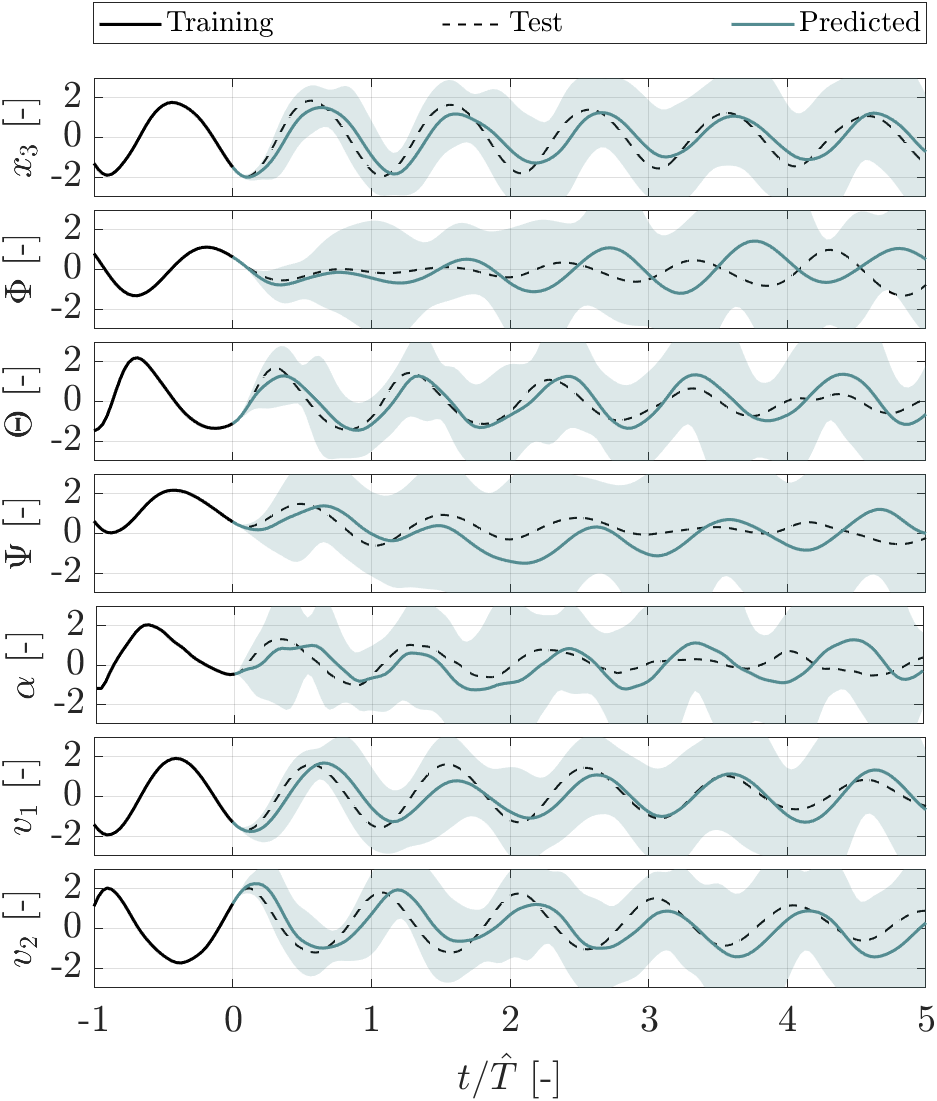}
        \caption{} \label{fig:bfore2-tempest}    
    \end{subfigure}
    \begin{subfigure}[b]{0.32\linewidth}   
        \includegraphics[width=\linewidth]{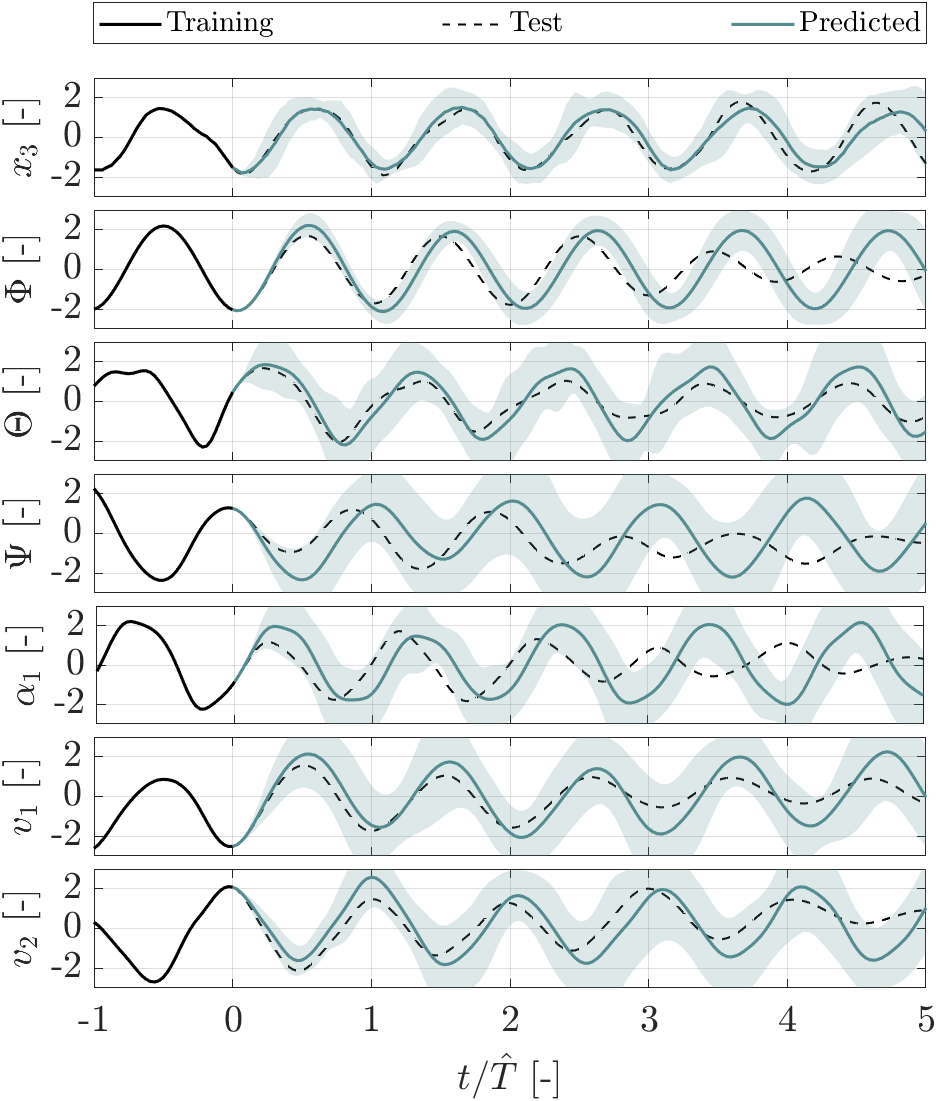}
        \caption{} \label{fig:bfore2-shipmo3d}    
    \end{subfigure}
\caption{\textit{Nowcasting} prediction of random time series by Hankel-DMD (best hyperparameters) and Bayesian Hankel-DMD. CFDShip-Iowa: (\subref{fig:fore2-iowa}) and (\subref{fig:bfore2-iowa}); TEMPEST: (\subref{fig:fore2-tempest}) and (\subref{fig:bfore2-tempest}); ShipMo3D: (\subref{fig:fore2-shipmo3d}) and (\subref{fig:bfore2-shipmo3d}).} \label{fig:fore2}
\end{figure}

\Cref{fig:fore1-iowa,fig:fore1-tempest,fig:fore1-shipmo3d} show the prediction generated using the Hankel-DMD \textit{nowcasting} for random time series from CFDShip-Iowa, TEMPEST, and ShipMo3D simulations, respectively.
The training (past) time histories are depicted in black, the predicted (future) time histories are in blue, and the true (future) time histories are presented with a dashed black line.
The forecasting are obtained using the hyperparameters pair that yielded the lowest average NRMSE error among the tested configurations, \textit{i.e.}, the best ones whose error metrics are reported in \cref{tab:det_an1}.
The depicted time series are indicative of the large subset of \textit{successful} predictions, showing excellent agreement in both phase and amplitude throughout the entire $5\hat{T}$ prediction time frame.
However, some less accurate predictions are also present in the set. \Cref{fig:fore2-iowa,fig:fore2-tempest,fig:fore2-shipmo3d} show predictions representative of the worst performances for the same hyperparameter settings of \cref{fig:fore1}.

\subsection{Bayesian algorithm}
The Bayesian extension of the algorithm is obtained by integrating the insights on the hyperparameters derived from the deterministic analysis, identifying the most promising range for the hyperparameters in terms of prediction accuracy.
The training time history length is treated as a probabilistic variable, uniformly distributed between 1 and 5 encounter wave periods, $l_{tr}/{\hat T} \sim \mathcal{U}(1,5)$. 
Additionally, for each realization of $l_{tr}$, $l_{d}$ is also considered a probabilistic variable, uniformly distributed within the interval $l_{d} \sim \mathcal{U}\left(\frac{1}{2}l_{tr},\frac{3 }{4}l_{tr}\right)$ 
(the actual $n_d$ is taken as the corresponding integer part).

\Cref{fig:bfore1-iowa,fig:bfore1-tempest,fig:bfore1-shipmo3d} and \cref{fig:bfore2-iowa,fig:bfore2-tempest,fig:bfore2-shipmo3d}
show the predictions obtained by the Bayesian extension for the same test sequences as shown in 
\cref{fig:fore1-iowa,fig:fore1-tempest,fig:fore1-shipmo3d}
\cref{fig:fore2-iowa,fig:fore2-tempest,fig:fore2-shipmo3d}
respectively. 
Chebyshev’s inequality is used for the shaded area representing uncertainty, with a coverage factor equal to 2 (88.89\% confidence interval). 

The enhancement in forecasting time series using the expected value of the Bayesian prediction compared to the deterministic approach is evident, in particular comparing \cref{fig:fore2-iowa,fig:fore2-tempest,fig:fore2-shipmo3d}, which was the reference for the \textit{less accurate} deterministic predictions, with \cref{fig:bfore2-iowa,fig:bfore2-tempest,fig:bfore2-shipmo3d}.
The expected prediction line recovers a satisfying accuracy level throughout the entire prediction window, significantly reducing the discrepancy with the true signals.

Interestingly, the uncertainty of the Bayesian prediction remains low for the histories in \cref{fig:fore1}, for which the deterministic algorithm yielded satisfying results. Conversely, it increases noticeably for the time histories shown in \cref{fig:fore2}.
The standard deviation of the forecast appears directly correlated with the accuracy of the prediction of the time series at hand, providing a useful tool to infer the reliability of the ROM.

\begin{figure}[ht!]
    \centering
    \includegraphics[width=\linewidth]{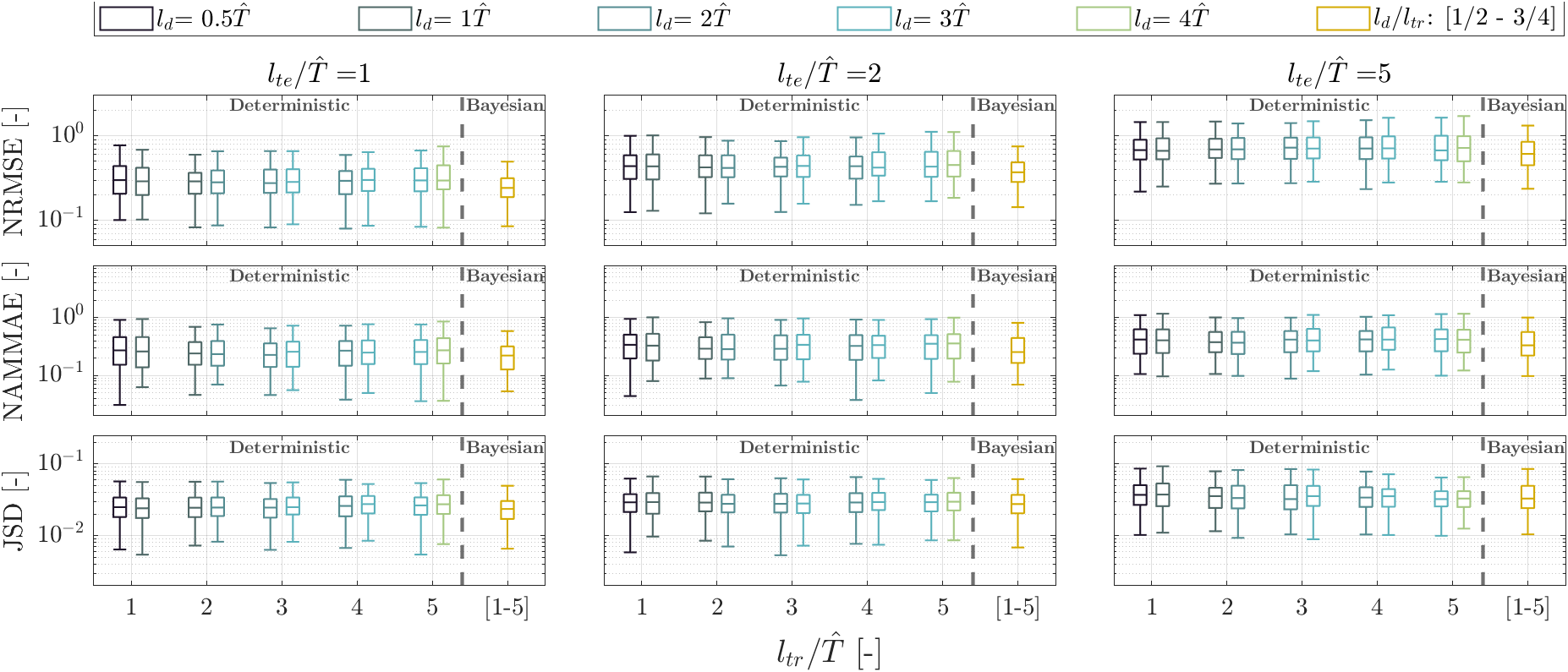}
    \caption{Error metrics from \textit{nowcasting} statistical analysis, deterministic vs Bayesian forecasting algorithm, CFDShip-Iowa data.}
    \label{fig:bstat-cfdship}
\end{figure}

\begin{figure}[ht!]
    \centering
    \includegraphics[width=\linewidth]{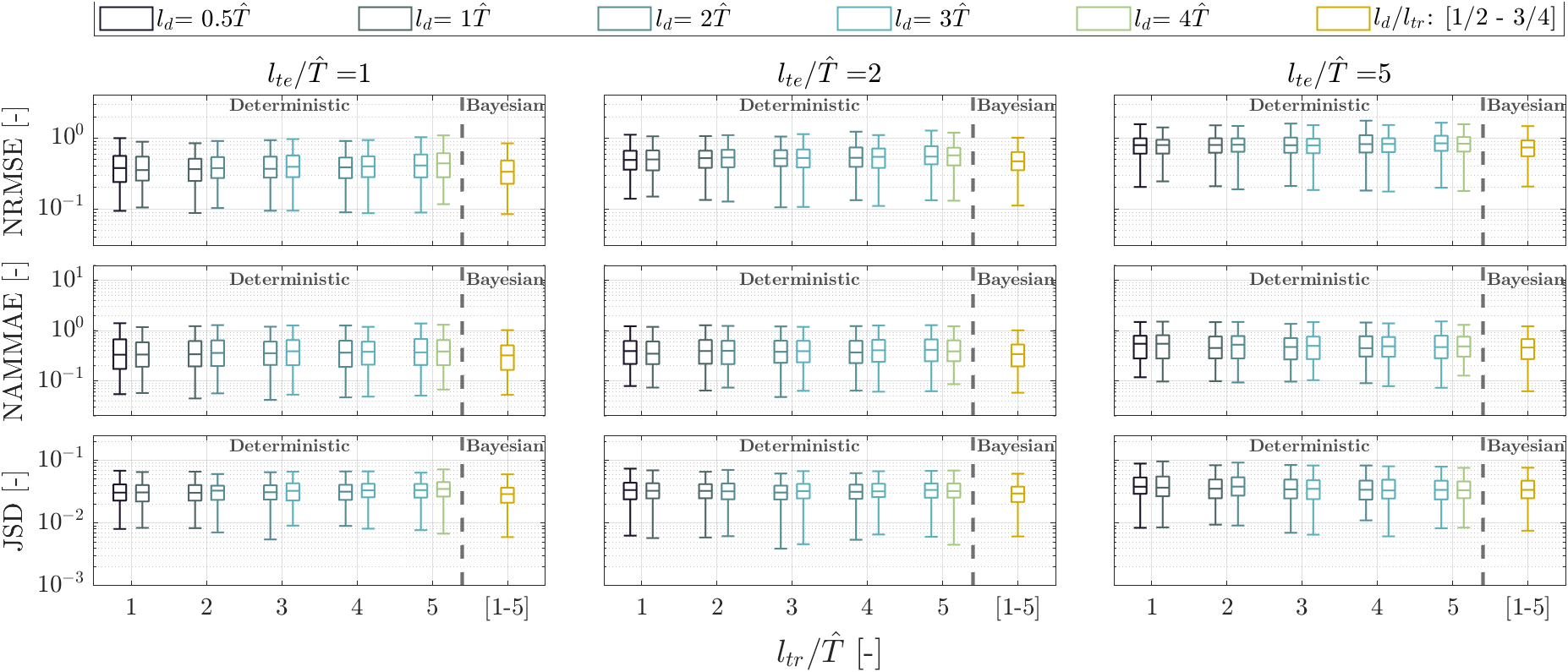}
    \caption{Error metrics from \textit{nowcasting} statistical analysis, deterministic vs Bayesian forecasting algorithm, TEMPEST data.}
    \label{fig:bstat-tempest}
\end{figure}

\begin{figure}[ht!]
    \centering
    \includegraphics[width=\linewidth]{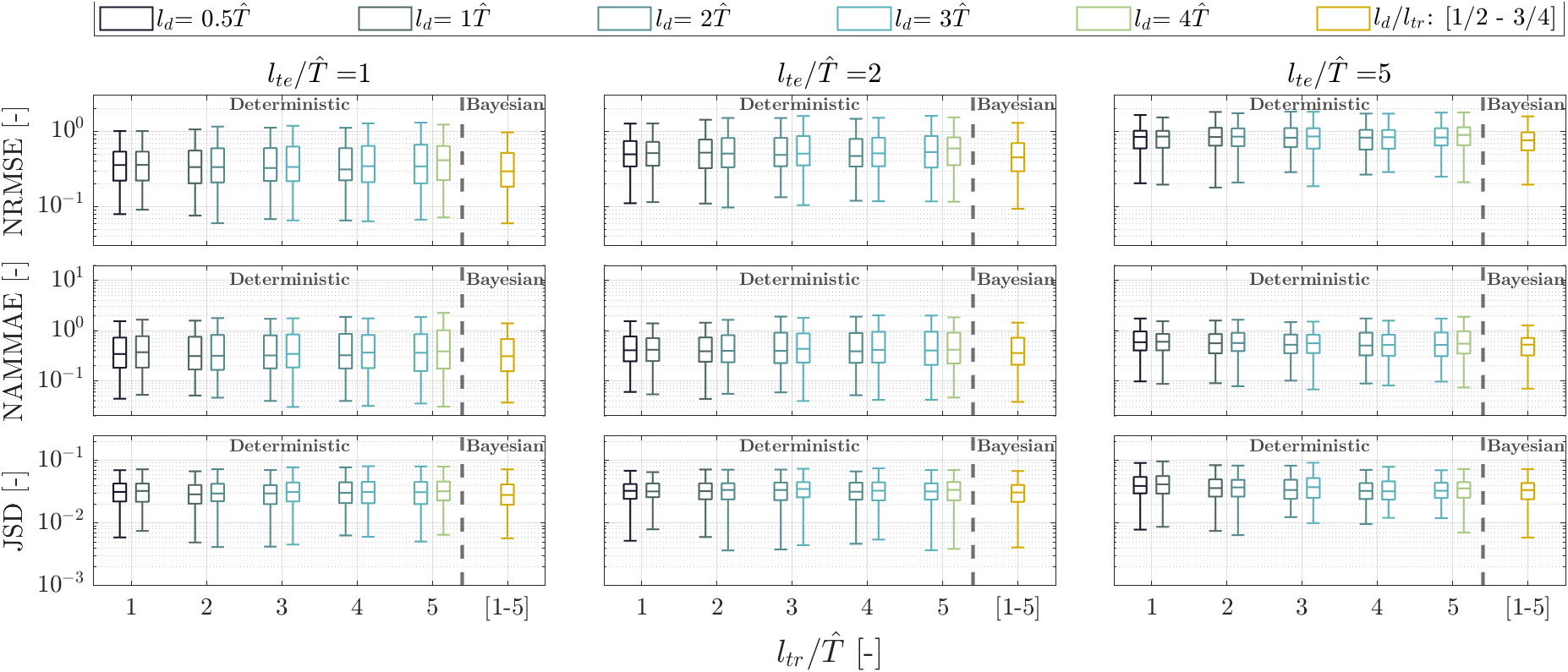}
    \caption{Error metrics from \textit{nowcasting} statistical analysis, deterministic vs Bayesian forecasting algorithm, ShipMo3D data.}
    \label{fig:bstat-shipmo}
\end{figure}

To systematically evaluate the performance of the Bayesian Hankel-DMD \textit{nowcasting}, we replicated the statistical analyses conducted with the deterministic algorithm using the same set of 250 random time series.
\Cref{fig:bstat-cfdship,fig:bstat-tempest,fig:bstat-shipmo} present a comparison of the metrics obtained from the deterministic and Bayesian DMD. 
The Bayesian algorithm consistently outperforms the deterministic one across all three metrics. This improvement is most pronounced for the shortest prediction window and diminishes for longer time horizons.
The improvement in the accuracy prediction obtained by the Bayesian algorithm is also confirmed by \cref{tab:det_an1,tab:det_an2,tab:det_an3}, where a comparison with the best deterministic cases is provided for the three metrics, showing a solid improvement in the average values and a reduction in the standard deviation of the performances for the great majority of the cases.

An important aspect of the proposed algorithm is its execution time. 
A statistical analysis is performed for the hyperparameter configuration involving the largest $\mathbf{A}$ matrix, $l_{tr}/\hat{T} = 5$ and $l_d/\hat{T} = 5$, repeating fifty times the run of a MATLAB 2023a implementation on a laptop mounting a $12^{th}$ Gen Intel(R) Core(TM) i5-1235U. 
The time needed to obtain the prediction, once the data are organized in the $\widehat{\mathbf{X}}$ and $\widehat{\mathbf{X}}'$ matrices, has been comprised in the range $t = [0.04s, 0.0717s]$, with average value $\mu_t = 0.0523s$, and standard deviation $\sigma_t = 0.0064s$.
The algorithm is hence capable of real-time forecasting, that can be continuously updated with incoming data from the physical twin. 
This holds also for the Bayesian extension of the algorithm, which requires multiple estimations with different hyperparameter configurations to build the stochastic prediction: the different runs can be performed in parallel being completely independent of each other, keeping the required time compatible with real-time executions.

\section{Conclusion}\label{s:concl}

This study proposed a Bayesian extension of the Hankel-DMD method for ship motions nowcasting, aimed at the development of digital twins for naval vessels operating in waves.
It addresses the need in this context for accurate and real-time predictions of ship performances to improve maritime operations' safety and efficiency.

This method builds a data-driven reduced-order model, using a small amount of data, for the prediction of ship motions in waves and continuously updates it with the incoming data from the physical system.
The approach has been tested on CFD simulated data for the course-keeping of the 5415M model in beam-quartering irregular waves and sea state 7 from three different solvers, showing the ability to forecast the system state in the near future, up to five wave encounter periods, with good accuracy. 

The Bayesian approach is developed from its deterministic version considering the method's hyperparameters as stochastic variables. Their ranges of variation are obtained after a statistical analysis of their effect on the prediction accuracy, based on three evaluation metrics covering different aspects of the time histories.

The results demonstrate that the Bayesian Hankel-DMD significantly improves the accuracy of the predictions compared to the deterministic method considering all the metrics. In addition, the Bayesian approach also enhances the predictions with uncertainty quantification, providing insights into their confidence level. The analysis showed a promising correlation between accuracy and uncertainty. This characteristic is crucial for digital twinning applications to inform high-regret decisions in the maritime environment.

The proposed nowcasting strategy is computationally very efficient, capable of providing a deterministic prediction in the order of a few hundredths of a second. The evaluation of a Bayesian forecast is an embarrassingly parallel task, suggesting that the same computational time can be achieved also for a stochastic prediction.

Future efforts will be devoted to considering different operational and environmental conditions, aiming to assess and improve the robustness of the DMD-based forecasting methods, also in transient maneuvering in waves situations. Methodological advancements including the use of other DMD extensions (\textit{e.g.}, DMD with control, see \cite{proctor2016dynamic}), or the combination with artificial neural networks approaches \cite{Diez2024}, will be explored to extend the capabilities of the forecasting method, overcoming some limitations that may arise in the use of the current approach in such challenging applications.

\section*{Acknowledgments}
The authors thank the financial support of the US Office of Naval Research, NICOP Grant N62909-21-1-2042, “Improving Knowledge, Prediction, and Forecasting of Ships in Waves via Hybrid Machine Learning Methods” (FORWARD), 
and of the Italian Ministry of University and Research through the National Recovery and Resilience Plan (PNRR), CN00000023- CUP B43C22000440001, “Sustainable Mobility Center” (CNMS), Spoke 3 “Waterways.” 
The authors are also grateful to NATO Science and Technology Organization, Applied Vehicle Technology task group 
AVT-351 (“Enhanced Computational Performance and Stability \& Control Prediction for NATO Military Vehicles”) for the fruitful collaboration through the years.
 
\section*{CRediT authorship contribution statement}
\textbf{Giorgio Palma:} Conceptualization, Methodology, Software, Validation, Investigation, Formal Analysis, Writing - Original Draft, Visualization.
\textbf{Andrea Serani:} Data Curation, Resources, Writing - Review \& Editing.
\textbf{Kevin McTaggart:} Data Curation, Resources, Writing - Review \& Editing.
\textbf{Shawn Aram:} Data Curation, Resources, Writing - Review \& Editing.
\textbf{David W. Wundrow:} Data Curation, Resources.
\textbf{David Drazen:} Data Curation, Resources, Writing - Review \& Editing.
\textbf{Matteo Diez:} Conceptualization, Methodology, Resources, Writing - Review \& Editing, Supervision, Funding acquisition.


\bibliographystyle{unsrt}  
\bibliography{biblio}  

\end{document}